\documentclass{article}
\usepackage[utf8]{inputenc}
\usepackage[left=3cm, right=3cm, top=2cm]{geometry}
\usepackage{amssymb,amsmath}
\usepackage{bm}
\usepackage{sistyle}
\usepackage{caption}
\usepackage{acronym}
\usepackage[algoruled]{algorithm2e}
\usepackage{graphicx}
\graphicspath{ {figures/} }
\usepackage{afterpage}
\usepackage[table]{xcolor}
\usepackage[colorlinks=true, allcolors=blue]{hyperref}
\usepackage{url}
\usepackage[capitalise]{cleveref}
\usepackage[retainorgcmds]{IEEEtrantools}
\usepackage[textsize=tiny]{todonotes}
\usepackage{natbib}
\usepackage{lineno}
\usepackage{subfig}
\usepackage[font=small, labelfont=bf]{caption}

\usepackage{titling}
\pretitle{\begin{center}\LARGE\sffamily\bfseries}
\posttitle{\par\end{center}\vskip 0.5em}
\preauthor{\begin{center}\normalsize
\lineskip 0.5em%
\begin{tabular}[t]{c}}
\postauthor{\end{tabular}\par\end{center}}
\predate{\begin{center}}
\postdate{\par\end{center}}
\newif\iffinal
\finaltrue
\iffinal
    \newenvironment{preview}{}{}
\else
    \usepackage[tightpage,active]{preview}
\fi

\usepackage{minitoc}

\newacro{ADHD}{Attention Deficit Hyperactivity Disorder}
\newacro{FDR}{False Discovery Rate}
\newacro{AUC}{Area Under the Curve}
\newacro{NLP}{natural language processing}
\newacro{BOW}{Bag-Of-Words}
\newacro{ROC}{Receiver Operating Characteristic}
\newacro{i.i.d.}{independent identically distributed}
\newacro{SVM}{Support Vector Machine}
\newacro{MNI}{Montreal Neurological Institute}
\newacro{GCV}{Generalized Cross-Validation}
\newacro{PET}{Positron Emission Tomography}
\newacro{MRI}{Magnetic Resonance Imaging}
\newacro{MKDA}{Multilevel Kernel Density Analysis}
\newacro{CBMA}{Coordinate-Based Meta-Analysis}
\newacro{IBMA}{Coordinate-Based Meta-Analysis}
\newacro{ALE}{Activation Likelihood Estimate}
\newacro{FWHM}{Full Width at Half Maximum}
\newacro{GCLDA}{Generalized Correspondence Latent Dirichlet Allocation}
\newacro{IBC}{Individual Brain Charting}
\newacro{JATS}{Journal Article Tag Suite}
\newacro{KDE}{Kernel Density Estimation}
\newacro{MeSH}{Medical Subject Headings}
\newacro{NMF}{Non-negative Matrix Factorization}
\newacro{ROI}{Region of Interest}
\newacro{RSVP}{Rapid-Serial-Visual-Presentation}
\newacro{TFIDF}{Term Frequency $\cdot$ Inverse Document Frequency}
\newacro{fMRI}{functional Magnetic Resonance Imaging}
\newacro{XSLT}{eXtensible Stylesheet Language Transformations}
\newacro{IQR}{InterQuartile Range}
\newacro{i.i.d}{independent identically distributed}

\newcommand{\para}{\paragraph{}}
\newcommand{\vs}{vs.\,}
\newcommand{\aka}{a.k.a.\,}
\newcommand{\ie}{i.e.\,}
\newcommand{\eg}{e.g.\,}

\newcommand{\X}{ \bm{X} }
\newcommand{\x}{ \bm{x} }
\newcommand{\Y}{ \bm{Y} }
\newcommand{\y}{ \bm{y} }

\newcommand{\B}{ \bm{B} }
\newcommand{\M}{ \bm{M} }
\newcommand{\m}{ \bm{m} }

\DeclareMathOperator*{\argmin}{argmin}

\DeclareMathOperator*{\Tr}{Tr}
\DeclareMathOperator*{\diag}{diag}
\DeclareMathOperator*{\Var}{Var}

\DeclareMathOperator*{\F}{F}
\DeclareMathOperator*{\tf}{tf}
\DeclareMathOperator*{\idf}{idf}
\DeclareMathOperator*{\df}{df}

\begin{document}

\title{NeuroQuery: comprehensive meta-analysis\\ of human brain mapping}

\author{Jérôme Dockès$^1$ \and Russell A. Poldrack$^2$ \and Romain Primet$^3$ \and
Hande Gözükan$^3$ \and
Tal Yarkoni$^4$ \and Fabian Suchanek$^5$ \and Bertrand Thirion$^1$ \and
Gaël Varoquaux$^{1,6}$}
\date{%
  $^1$Inria, CEA, Université Paris-Saclay \\
  $^2$Stanford University \\
  $^3$Inria \\
  $^4$University of Texas at Austin \\
  $^5$Télécom Paris, Institut Polytechnique de Paris\\
  $^6$Montréal Neurological Institute, McGill University
}

\maketitle
\begin{abstract}
Reaching a global view of brain organization requires assembling evidence
on widely different mental processes and mechanisms.
The variety of human neuroscience concepts and terminology poses a fundamental
challenge to relating brain imaging results across the scientific
literature.
Existing meta-analysis methods perform statistical tests on sets of
publications associated with a particular concept.
Thus, large-scale meta-analyses only tackle single terms that occur frequently.
We propose a new paradigm, focusing on prediction rather than inference.
Our multivariate model \emph{predicts} the spatial distribution of neurological
observations, given text describing an experiment, cognitive process, or disease.
This approach handles text of arbitrary length and terms that are too rare for
standard meta-analysis.
We capture the relationships and neural correlates of 7\,547 neuroscience terms
across 13\,459 neuroimaging publications.
The resulting meta-analytic tool, \href{https://neuroquery.org}{neuroquery.org},
can ground hypothesis generation and data-analysis priors on a comprehensive
view of published findings on the brain.

\end{abstract}

\doparttoc 
\faketableofcontents 

\section{Introduction: pushing the envelope of meta-analyses}

Each year, thousands of brain-imaging studies explore the links between brain
and behavior: more than 6\,000 publications a year contain the term
``neuroimaging'' on PubMed.
Finding consistent trends in the knowledge acquired across these studies is
crucial, as individual studies by themselves seldom have enough statistical
power to establish fully trustworthy results
\citep{button2013power,poldrack2017scanning}.
But compiling an answer to a specific question from this impressive number of
results is a daunting task. 
There are too many studies to manually 
collect and aggregate their findings. In addition, such a task is
fundamentally difficult due to
the many different aspects of behavior, as well as the diversity
of the protocols used to probe them.
%

Meta-analyses can give objective views of the field, to ground a review
article or a discussion of new results.
\ac{CBMA} methods \citep{laird2005brainmap,wager2007meta,eickhoff2009coordinate}
assess the consistency of results across studies,
comparing the observed spatial density of reported brain
stereotactic coordinates to the null hypothesis of a uniform
distribution. Automating \ac{CBMA}
methods across the literature, as 
in NeuroSynth \citep{yarkoni2011large}, enables large-scale analyses of
brain-imaging studies, giving excellent statistical power.
Existing meta-analysis methods focus on identifying effects reported
consistently across the literature, to
distinguish true discoveries from noise and artifacts.
However, they can only address neuroscience concepts that are easy to define.
Choosing
which studies to include in a meta-analysis can be challenging.
In principle,
studies can be manually annotated as carefully as one likes. However,
manual meta-analyses are not scalable, and the corresponding degrees of
freedom are difficult to control statistically.
In what follows, we focus on automated
meta-analysis.
To automate the selection of studies,
the common solution is to rely on terms present in publications.
But closely related terms can lead to markedly different meta-analyses
(\cref{fig:maps-related-to-calculation}).
The lack of a universally established vocabulary or ontology to
describe mental processes and disorders is a strong impediment to meta-analysis
\citep{poldrack2016brain}.
Indeed, only 30\% of the terms contained in a neuroscience ontology or
meta-analysis tool are common to another (see
\cref{table:voc-intersections}).
In addition,
studies are diverse in many ways: they investigate different mental processes,
using different terms to describe them, and different experimental
paradigms to probe them \citep{newell1973you}.
Yet, current meta-analysis approaches model all studies as asking 
the same question. They cannot model nuances across studies because they
rely on in-sample statistical inference and are not designed
to interpolate between studies that address related but different
questions, or make predictions for unseen combinations of mental processes.
A consequence is that,
as we will show, their results are harder to control outside of
well-defined and frequently-studied psychological
concepts.

\begin{table}[t!]%
\def\cell#1{\cellcolor{green!#1} #1\%}%
\center


{\sffamily\small
\begin{tabular}{lllllll}
\bfseries {\% of $\downarrow$} contained in $\rightarrow$ & \rlap{\rotatebox{45}{Cognitive Atlas}}\hspace*{3.5ex}\rlap{\rotatebox{45}{\small (895)}} &
\rlap{\rotatebox{45}{MeSH}}\hspace*{3.5ex}\rlap{\rotatebox{45}{\small (21287)}} &
\rlap{\rotatebox{45}{NeuroNames}}\hspace*{3.5ex}\rlap{\rotatebox{45}{\small (7146)}} &
\rlap{\rotatebox{45}{NIF}}\hspace*{3.5ex}\rlap{\rotatebox{45}{\small (6912)}} &
\rlap{\rotatebox{45}{NeuroSynth}}\hspace*{3.5ex}\rlap{\rotatebox{45}{\small (1310)}} &
\rlap{\rotatebox{45}{\textbf{NeuroQuery}}}\hspace*{3.5ex}\rlap{\rotatebox{45}{\small (7547)}} \\
\hline
Cognitive Atlas &             \cell{100} &     \cell{14} &           \cell{0} &    \cell{3} &          \cell{14} &          \cell{68} \\
MeSH            &               \cell{1} &    \cell{100} &           \cell{3} &    \cell{4} &           \cell{1} &           \cell{9} \\
NeuroNames      &               \cell{0} &      \cell{9} &         \cell{100} &   \cell{29} &           \cell{1} &          \cell{10} \\
NIF             &               \cell{0} &     \cell{12} &          \cell{30} &  \cell{100} &           \cell{1} &          \cell{10} \\
NeuroSynth      &               \cell{9} &     \cell{14} &           \cell{5} &    \cell{5} &         \cell{100} &          \cell{98} \\
\textbf{NeuroQuery}      &               \cell{8} &     \cell{25} &           \cell{9} &    \cell{9} &          \cell{17} &         \cell{100} \\
\end{tabular}%
}
  \caption{
    \textbf{Diversity of vocabularies}: there is no established
    lexicon of neuroscience, even in hand-curated reference vocabularies, as visible across
    CognitiveAtlas \citep{poldrack2016brain}, MeSH
    \citep{lipscomb2000medical}, NeuroNames \citep{bowden1995neuronames}, NIF
    \citep{gardner2008neuroscience}, and NeuroSynth
    \citep{yarkoni2011large}.
    %
    Our dataset, NeuroQuery, contains all the terms from the other vocabularies
    that occur in more than 5 out of 10\,000 articles.
    ``MeSH'' corresponds to the branches of PubMed's MEdical Subject Headings
    related to neurology, psychology, or neuroanatomy (see
    \cref{parts-of-mesh-used}).
    Many MeSH terms are hardly or never used in practice -- \eg variants of
    multi-term expressions with permuted word order such as ``Dementia,
    Frontotemporal'', and are therefore not included in NeuroQuery's vocabulary. }
  \label{table:voc-intersections}
\end{table}

\para
Currently, an automated meta-analysis cannot cover all studies that report a
particular functional contrast (contrasting mental conditions to isolate a
mental process, \citet{poldrack2011handbook}). Indeed, we lack the tools to
parse the text in articles and reliably identify those that relate to equivalent
or very similar contrasts.
As an example, consider a study of the neural support of
translating orthography to phonology, probed with visual stimuli by
\citet{pinho2018individual}. The results of this study build upon an
experimental contrast labeled by the authors as ``Read pseudo-words \vs
consonant strings'', shown in \cref{fig:roi-ibc}.
%
Given this description, what prior hypotheses arise from the literature
for this contrast? Conversely, given the statistical map resulting from 
the experiment, how can one compare it with previous reports on similar tasks?
For these questions, meta-analysis seems the tool of choice.
Yet, the current meta-analytic paradigm requires the practitioner to select a
set of studies that are included in the meta-analysis.
In this case, which studies from the literature should be included?
Even with a corpus of 14\,000 full-text articles, selection based on simple
pattern matching --as with NeuroSynth-- falls short. Indeed, only 29 studies
contain all 5 words from the contrast description, which leads to a noisy and
under-powered meta-analytic map (\cref{fig:roi-ibc}).
To avoid relying on the contrast name, which can be seen as too short and terse,
one could do a meta-analysis based on the page-long task
description\footnote{\url{https://project.inria.fr/IBC/files/2019/03/documentation.pdf}}.
However, that would require combining even more terms, which precludes selecting
studies that contain all of them.
A more manual selection may help to identify relevant studies, but it
is far more difficult and time-consuming.
Moreover, some concepts of interest may not have been investigated by themselves,
or only in very few studies: rare diseases, or tasks
involving a combination of mental processes that have not been studied
together. For instance, there is evidence of
agnosia in Huntington's disease \citep{sitek2014unawareness}, but it has
not been studied with brain imaging. To compile a brain map from the literature
for such queries, it is necessary to interpolate between studies only partly
related to the query.
Standard meta-analytic methods lack an automatic way to measure the
relevance of studies to a question, and to interpolate between them. This
prevents them from answering new questions, or questions that cannot be
formulated simply.

Many of the constraints of standard meta-analysis arise from the
necessity to define an \emph{in-sample} test on a given set of studies.
Here, we propose a new kind of meta-analysis, that
focuses on \emph{out-of-sample} prediction rather than hypothesis testing.
The focus shifts from establishing consensus for a particular subject of
study to
building multivariate mappings from mental diseases and psychological concepts to anatomical structures in the brain.
This approach is complementary to classic meta-analysis 
methods such as \ac{ALE} \citep{laird2005brainmap}, \ac{MKDA}
\citep{wager2007meta} or NeuroSynth \citep{yarkoni2011large}: these
perform statistical tests to evaluate
trustworthiness of results from past studies, while our framework
predicts, based on the description of an experiment or subject of study,
which brain regions are most likely to be observed in a study.
We introduce a new meta-analysis tool, NeuroQuery, that predicts the neural
correlates of neuroscience concepts -- related to behavior, diseases, or anatomy.
To do so, it considers terms not in isolation, but in a dynamic,
contextually-informed way that allows for mutual interactions.
A predictive framework enables maps to be generated by generalizing from
terms that are well studied (``faces'') to those that are less well studied
and inaccessible to traditional meta-analyses (``prosopagnosia'').
As a result, NeuroQuery produces high-quality brain maps for concepts
studied infrequently in the literature and for a larger
class of queries than existing tools -- including, e.g., free text descriptions
of a hypothetical experiment.
These brain maps predict well the spatial distribution of findings and
thus form good grounds to generate regions of interest or interpret
results for studies of infrequent terms such as prosopagnosia.
Yet, unlike with conventional meta-analysis, they
do not control a voxel-level null hypothesis, hence are less suited to
asserting that a particular area
is activated in studies, e.g. of prosopagnosia.

\para
Our approach, NeuroQuery, assembles results from the literature into a brain
map using an arbitrary query with words from our vocabulary
of 7\,547 neuroscience terms.
NeuroQuery uses a multivariate model of the statistical link between multiple
terms and corresponding brain locations. It is
fitted using supervised machine learning on 13\,459
full-text publications.
Thus, it learns to weight and combine terms to predict the brain locations most
likely to be reported in a study. It can predict a brain map given any
combination of terms
related to neuroscience -- not only single words, but also detailed
descriptions, abstracts, or full papers.
With an extensive comparison to published studies, we show in 
\cref{encoding-papers-experiment} that it indeed approximates well
results of actual experimental data collection.
NeuroQuery also models the semantic relations that underlie the vocabulary of neuroscience. 
Using techniques from \acl{NLP}, NeuroQuery infers
semantic similarities across terms used in the literature.
Thus, it makes better use of the available information, and can recover
biologically plausible brain maps where other automated methods lack statistical
power, for example with terms that are used in few studies, as shown in
\cref{results-rare-concepts}.
This semantic model also makes NeuroQuery less sensitive to small variations in
terminology (\cref{fig:maps-related-to-calculation}).
Finally, the semantic similarities captured by NeuroQuery can help researchers 
navigate related neuroscience concepts while exploring their
associations with brain activity.
NeuroQuery extends the scope of standard meta-analysis, as it
extracts from the literature a comprehensive
statistical summary of evidence accumulated by neuroimaging research.
It can be used to explore the domain knowledge across sub-fields,
generate new hypotheses, and construct quantitative priors or regions of
interest for future studies, or put in perspective results of an experiment.
%
NeuroQuery is easily usable online, at
\href{https://neuroquery.org}{neuroquery.org}, and the data and source code can
be freely downloaded.
We start by briefly describing the statistical model behind NeuroQuery in
\cref{overview-of-the-neuroquery-model}, then illustrate its usage
(\cref{illustration}) and show that it can map new combinations of
concepts in \cref{new-combinations-of-concepts}. In \cref{results-rare-concepts} and
\ref{encoding-papers-experiment}, we conduct a thorough qualitative and
quantitative assessment of the new possibilities it offers, before a
discussion and conclusion.

\section{Results: The NeuroQuery tool and what it can do.}

\subsection{Overview of the NeuroQuery model.}\label{overview-of-the-neuroquery-model}

\begin{figure}[b]
  \centering
  \iffinal
    \includegraphics[height=.41\textwidth]{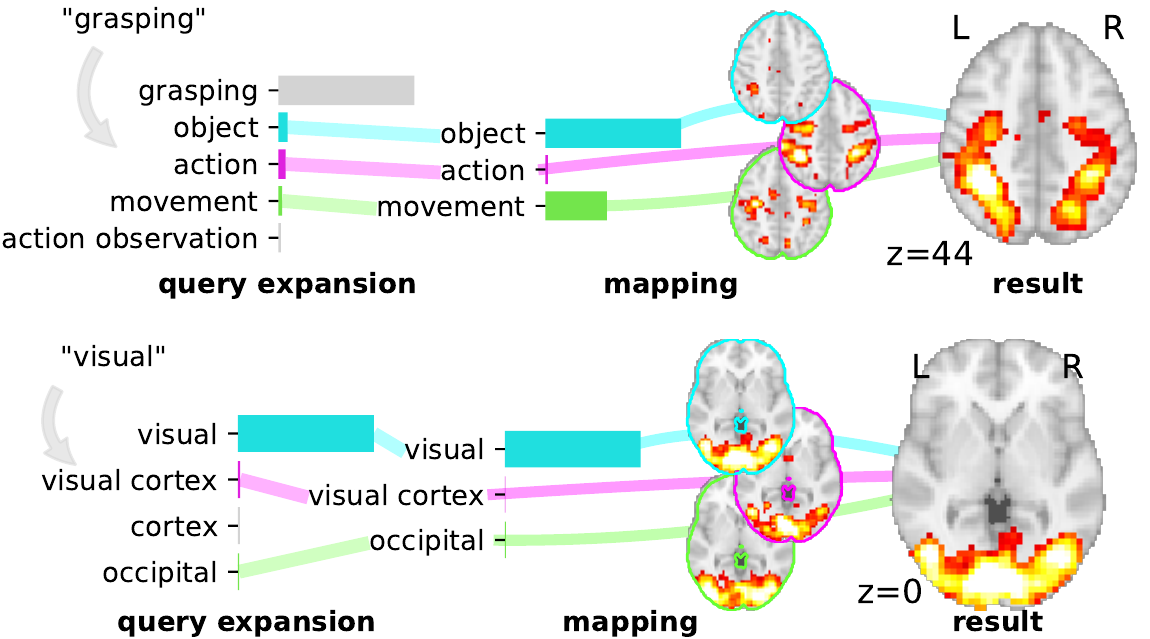}
  \fi%
  \caption{\textbf{Overview of the NeuroQuery model}: two examples of how
association maps are constructed for the terms ``grasping'' and ``visual''. The
query is expanded by adding weights to related terms. The resulting vector is
projected on the subspace spanned by the smaller vocabulary selected during
supervised feature selection. Those well-encoded terms are shown in color.
Finally, it is mapped onto the brain space through the regression model. When a
word (e.g., ``visual'') has a strong association with brain activity and is
selected as a regressor, the smoothing has limited effect.
\textbf{Details:} the first bar plot shows the semantic similarities of
neighboring terms with the query. It represents the smoothed \ac{TFIDF} vector.
Terms that are not used as features for the supervised regression are shown in
gray. The second bar plot shows the similarities of selected terms, rescaled by
the norms of the corresponding regression coefficient maps. It represents the
relative contribution of each term in the final prediction. The coefficient maps
associated with individual terms are shown next to the bar plot. These maps are
combined linearly to produce the prediction shown on the right. }
    \label{fig:overview}
\end{figure}

NeuroQuery is a statistical model that identifies brain regions
related to an arbitrary text query -- a single term, a few keywords, or a longer text.
It is built on a controlled vocabulary of neuroscience terms and a
large corpus containing the full text of neuroimaging publications and
the coordinates that they report.
The main components of the NeuroQuery model are an estimate of the relatedness
of terms in the vocabulary, derived from co-occurrence statistics, and a
regression model that links term occurrences to neural activations using
supervised machine learning techniques.
To generate a brain map, NeuroQuery first uses the estimated semantic
associations to map the query onto a set of keywords that can be reliably associated
with brain regions.
Then, it transforms the resulting representation into a brain map using
a linear regression model (\cref{fig:overview}).
This model can thus be understood as a reduced rank regression, where the
low-dimensional representation is a distribution of weights over keywords
selected for their strong link with brain activity.
We emphasize the fact that NeuroQuery is a \emph{predictive} model. The maps it
outputs are predictions of the likelihood of observation brain location
(rescaled by their standard deviation). They do not have the same meaning as
\ac{ALE}, \ac{MKDA} or NeuroSynth maps as they do not show a voxel-level test
statistic.
In this section we describe our neuroscience corpus and how we use it to
estimate semantic relations, select keywords, and map them onto brain
activations.

\para
NeuroQuery relies on a corpus of 13\,459 full-text neuroimaging publications,
described in \cref{section:building-neuroquery-data}. This corpus is by far the
largest of its kind; the NeuroSynth corpus contains a similar number of
documents, but uses only the article abstracts, and not the full article texts.
We represent the text of a document with the (weighted) occurrence frequencies
of each phrase from a fixed vocabulary, i.e., \acf{TFIDF} features
\citep{salton1988term}.
This vocabulary is built from the union of terms from several ontologies
(shown in
\cref{table:voc-intersections}) and labels from 12 anatomical atlases (listed in
\cref{table:atlases-used} in \cref{atlases-used}). It
comprises 7\,547 terms or phrases related to neuroscience that occur in at least
0.05\% of publications.
%
We automatically extract 418\,772 peak activations coordinates from
publications, and transform them to brain maps with a kernel density
estimator. Coordinate extraction is discussed and evaluated in
\cref{subsection:coordinate-extraction}.
This preprocessing step thus yields, for each article:
  its representation in term frequency space (a \ac{TFIDF} vector), and
  a brain map representing the estimated density of activations for this study.
The corresponding data is also openly available online.

\para
The first step of the NeuroQuery pipeline is a semantic smoothing of the
term-frequency representations.
Many expressions are challenging for existing automated meta-analysis frameworks,
because they are too rare, polysemic, or have a low correlation with brain
activity. Rare words are problematic because peak activation coordinates
are a very weak signal: from each article we extract little information about
the associated brain activity. Therefore existing frameworks rely on the occurrence of a term
in hundreds of studies in order to detect a pattern in peak activations.
Term co-occurrences, on the other hand, are more consistent and reliable, and
capture semantic relationships \citep{turney2010frequency}. The strength
of these relationships encode semantic proximity, from very strong for
synonyms that occur in statistically identical contexts, to weaker
for different yet related mental processes that are often studied
one opposed to the other. Using them helps meta analysis:
it would require hundreds of studies to detect a pattern in locations
reported for
``aphasia'', \eg in lesion studies. But with the text of a few publications we notice that it often
appears close to ``language'', which is indeed a related mental process. By leveraging this information,
NeuroQuery recovers maps for terms that are too rare to be mapped
reliably with standard automated meta-analysis.
Using \ac{NMF}, we compute a low-rank approximation of word co-occurrences (the
covariance of the \ac{TFIDF} features), and obtain a denoised semantic
relatedness matrix (details are provided in
\cref{subsection:smoothing-details}). These word associations guide the encoding
of rare or difficult terms into brain maps. They can also be used to explore
related neuroscience concepts when using the NeuroQuery tool.

\para The second step from a text query to a brain map is
NeuroQuery's text-to-brain encoding model.
When analyzing the literature, we fit a linear regression to reliably
map text onto brain activations.
The intensity (across the peak density maps) of each voxel in the
brain is regressed on the \ac{TFIDF} descriptors of documents.
This model is an additive one across the term occurrences, as opposed to logical
operations traditionally used to select studies for meta-analysis. It results in
higher predictive power (\cref{document-frequencies}).

%
One challenge is that \ac{TFIDF} representations
are sparse and high-dimensional. We use a reweighted ridge regression and
feature selection procedure (described in \cref{reweighted-ridge-regression}) to
prevent uninformative terms such as ``magnetoencephalography'' from degrading
performance. This procedure automatically selects around 200 keywords that
display a strong statistical link with brain activity and adapts the
regularization applied to each feature.
Indeed, mapping too many terms (covariates) without appropriate regularization
would degrade the regression performance due to multicolinearity.

\para
To make a prediction, NeuroQuery combines semantic smoothing and linear
regression of brain activations. To encode a new document or query, the text is
expanded, or smoothed, by adding weight to related terms using the semantic
similarity matrix. The resulting smoothed representation is projected onto the
reduced vocabulary of selected keywords, then mapped onto the brain through the
linear regression coefficients (\cref{fig:overview}).
The rank of this linear model is therefore the size of the restricted
vocabulary that was found to be reliably mapped to the brain.
Compared with other latent factor models, this 2-layer linear model is easily
interpretable, as each dimension (both of the input and the latent space) is
associated with a term from our vocabulary.
In addition, NeuroQuery uses an estimate of the voxel-level variance of
association (see methodological details in \cref{methodological-details}),
and reports a map of Z statistics. Note that this variance represents an
uncertainty around a \emph{prediction} for a \ac{TFIDF} representation of the
concept of interest, which is treated as a fixed quantity.
Therefore, the resulting map
cannot be thresholded to reject any simple null hypothesis. NeuroQuery
maps have a different meaning and different uses than standard meta-analysis
maps obtained e.g.\, with \ac{ALE}.
%

\begin{figure}[h!]
  \iffinal
    \centerline{%
	\includegraphics[width=.9\textwidth]{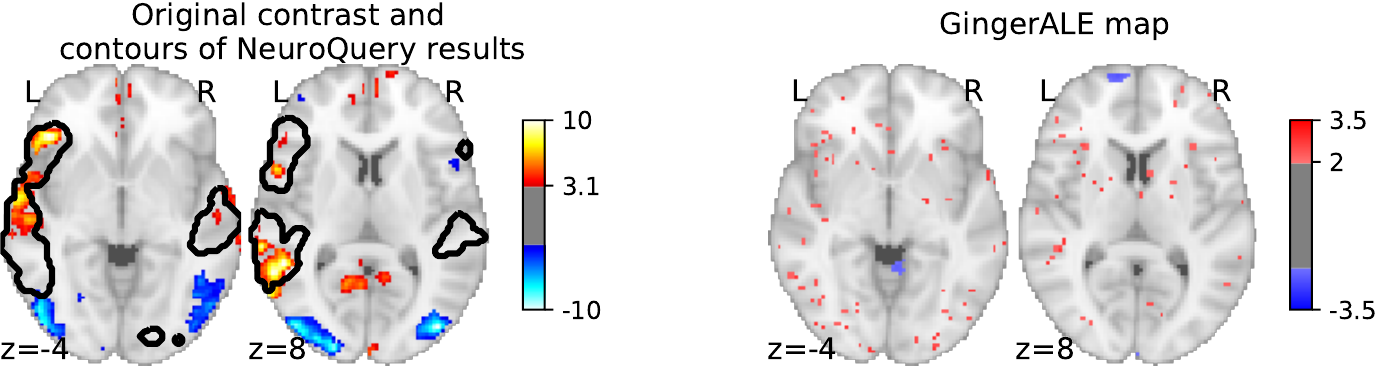}%
    }
  \fi%

\caption{\textbf{Illustration: studying the contrast ``Read pseudo words
\vs consonant strings''}.
\textbf{Left}: Group-level map from the IBC dataset for the contrast ``Read
pseudo-words \vs consonant strings'' and contour of NeuroQuery map obtained
from this query. The NeuroQuery map was obtained directly from the
contrast description in the dataset's documentation, without needing to
manually select studies for the meta-analysis nor convert this
description to a string pattern usable by existing automatic
meta-analysis tools.
The map from which the contour is drawn, as well as a NeuroQuery map for the
page-long description of the \ac{RSVP} language task, are shown in
\cref{section:details-ibc-experiment}, in
\cref{section:details-ibc-experiment}c and
\cref{section:details-ibc-experiment}d respectively.
\textbf{Right}: \ac{ALE}
map for 29 studies that contain all terms from the IBC contrast
description. The map was obtained with the GingerALE tool
\citep{eickhoff2009coordinate}. With only 29 matching studies,
\ac{ALE} lacks statistical power for this contrast description.}
\label{fig:roi-ibc}
\end{figure}

\subsection{Illustration: using NeuroQuery for post-hoc interpretation}
\label{illustration}

After running a \ac{fMRI} experiment, it is common to compare the
computed contrasts to what is known from the existing literature, and
even use prior knowledge to assess whether some activations are not
specific to the targeted mental process, but due to experimental
artifacts such as the stimulus modality.
It is also possible to introduce prior knowledge earlier in the study and
choose a \acf{ROI} before running the experiment.
This is usually done based on the expertise of the researcher, which is
hard to formalize and reproduce. With NeuroQuery, it is easy to capture
the domain knowledge and perform these comparisons or \ac{ROI} selections
in a principled way.
%

As an example, consider again the contrast from the \ac{RSVP} language task
\citep{pinho2018individual,humphries2006syntactic} in the \ac{IBC} dataset, shown
in \cref{fig:roi-ibc}. It is described as ``Read pseudo-words \vs consonant
strings''. We obtain a brain map from NeuroQuery by simply transforming the
contrast description, without any manual intervention, and compare both maps by
overlaying a contour of the NeuroQuery map on the actual \ac{IBC} group contrast
map.
We can also obtain a meta-analytic map for the whole \ac{RSVP} language task by
analyzing the free-text task description with NeuroQuery
(\cref{section:details-ibc-experiment}).

\subsection{NeuroQuery can map new combinations of
concepts}\label{new-combinations-of-concepts}

\begin{figure}[t]
  \iffinal
    \includegraphics[width=\textwidth]{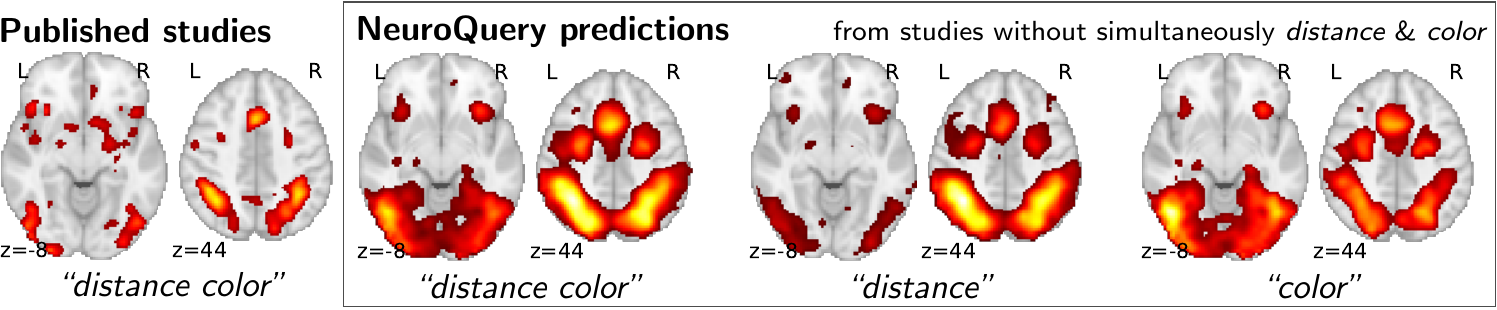}
  \fi%
\caption{\textbf{Mapping an unseen combination of terms}
\textsl{Left}
The difference in the spatial distribution of findings reported in
studies that contains both \textsl{``distance''} and \textsl{``color''}
($n=687$), and the rest of the studies.
-- 
\textsl{Right} Predictions of a NeuroQuery model fitted on the studies that
do not contain simultaneously both terms \textsl{``distance''}
and \textsl{``color''}.
\label{fig:unseen_combination}
}
\end{figure}

To study the predictions of NeuroQuery, we first demonstrate that it can
indeed give good brain maps on combinations of terms that have never been
studied together. For this, we leave out from our corpus of studies all
the publications that simultaneously mention two given terms, we fit a
NeuroQuery model on the resulting reduced corpus, and evaluate its
predictions on the left out publications, that did actually report these
terms together. \cref{fig:unseen_combination} shows an example of such an
experiment: excluding publications mentioning simultaneously ``distance''
and ``color''. The figure compares a simple meta analysis of the
combination of these two terms -- contrasting the left-out studies with
the remaining ones -- with the predictions of the model fitted excluding
studies that include the term conjunction. Qualitatively, the predicted maps comprise
all the brain structures visible in the simultaneous studies of
``distance'' and ``color'': on the one hand, the intra-parietal sulci,
the frontal eye fields, and the anterior cingulate / anterior insula
network associated with distance perception, and on the other hand, the
additional mid-level visual region around the approximate location of V4
associated with color perception. The extrapolation from two terms for
which the model has seen studies, ``distance'' and ``color'', to their
combination, for which the model has no data, is possible thanks to the
linear additive model, combining regression maps for ``distance'' and
``color''.

To assert that the good generalization to unseen pairs of terms is not
limited to the above pair, we apply
quantitative experiments of prediction quality (introduced later, in
\cref{encoding-papers-experiment}) to 1\,000 randomly-chosen pairs. We find
that measures of how well predictions match the literature decrease only
slightly for studies with terms already seen
together compared to studies with terms never seen jointly (details in
\cref{section:unseen-pairs-performance}). Finally, we gauge the quality
of the maps with a quantitative experiment mirroring the qualitative evaluation of
\cref{fig:unseen_combination}: for each of the 1\,000 pairs of terms, we
compute the Pearson correlation of the predicted map for the unseen
combination of
terms with the meta-analytic map obtained on the left-out studies. We find
a median correlation of 0.85 which shows that the excellent performance
observed on \cref{fig:unseen_combination} is not due to a specific choice
of terms.

\subsection{NeuroQuery can map rare or difficult concepts}\label{results-rare-concepts}

We now we compare the NeuroQuery model to
existing automated meta-analysis methods, investigate how it handles terms that are
challenging for the current state of the art, and quantitatively evaluate its
performance. We compare NeuroQuery with NeuroSynth
\citep{yarkoni2011large}, the best known automated meta-analytic tool, and
with \acf{GCLDA} \citep{rubin2017decoding}.
\ac{GCLDA} is an important baseline because it is the only multivariate
meta-analytic model to date. However, it produces maps with a
low spatial resolution because it models brain activations as a mixture of
Gaussians. Moreover, it takes several days to train and a dozen of seconds to
produce a map at test time, and is thus unsuitable to build an online and
responsive tool like NeuroSynth or NeuroQuery.

\begin{figure}[t!]
\hspace{-40pt}
\begin{minipage}{1.2\textwidth}
  \iffinal
    \includegraphics[width=\textwidth]{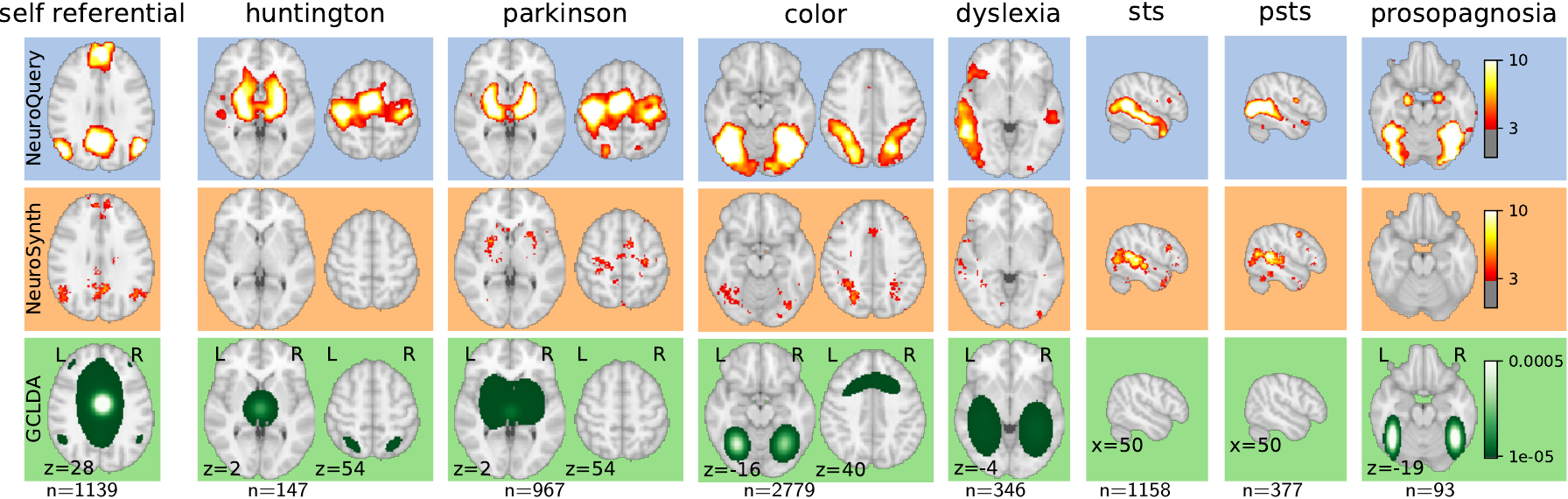}
  \fi%
  \caption{\textbf{Examples of maps obtained for a given term}, compared across
different large-scale meta-analysis frameworks.
``\ac{GCLDA}'' has low spatial resolution and produces
inaccurate maps for many terms. For relatively straightforward terms like
``psts'' (posterior superior temporal sulcus), NeuroSynth
and NeuroQuery give consistent results. For terms that are more rare or
difficult to map like ``dyslexia'', only NeuroQuery generates usable brain maps.}

  \label{fig:example-maps}
\end{minipage}
\end{figure}

\para
By combining term similarities and an additive encoding model, NeuroQuery can
accurately map rare or difficult terms for which standard meta-analysis lacks
statistical power, as visible on \cref{fig:example-maps}.

\begin{figure}
  \iffinal
    \includegraphics[width=\textwidth]{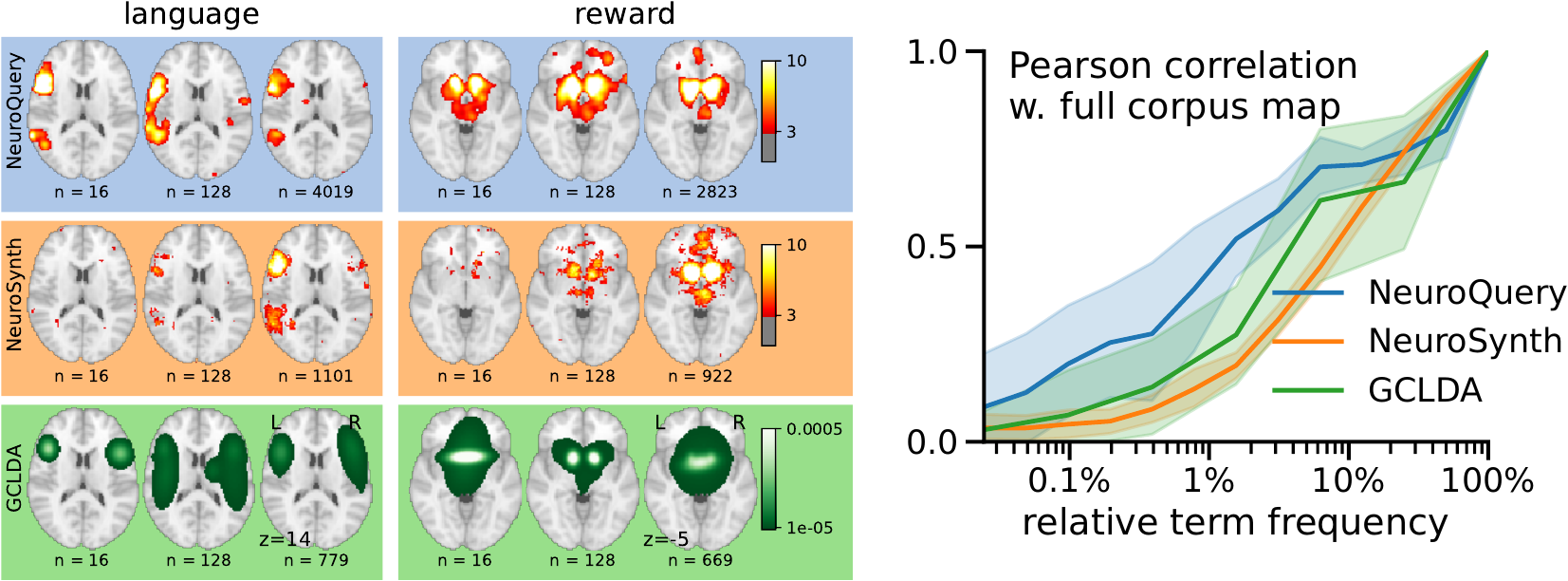}
  \fi%
\smallskip

\begin{minipage}{1.05\linewidth}
  \caption{
\textbf{Learning good maps from few studies.}
\textbf{left:} maps obtained from subsampled corpora, in which the encoded word
appears in 16 and 128 documents, and from the full corpus. NeuroQuery needs less
examples to learn a sensible brain map. NeuroSynth maps correspond to
NeuroSynth's Z scores for the ``association test'' from \url{neurosynth.org}.
NeuroSynth's ``posterior probability'' maps for these terms for the full corpus
are shown in \cref{fig:neurosynth-posterior-proba-maps}. Each tool is trained on
its own dataset, which is why the full-corpus occurrence counts differ.
\textbf{right:} convergence of maps toward their value for the full corpus, as the
number of occurrences increases. Averaged over 13 words: ``language'', ``auditory'', ``emotional'', ``hand'', ``face'', ``default mode'',
``putamen'', ``hippocampus'', ``reward'', ``spatial'', ``amygdala'',
``sentence'', ``memory''.
On average, NeuroQuery is closer to the full-corpus map. This confirms
quantitatively what we observe for the two examples ``language'' and ``reward''
on the left.
Note that here convergence is only measured with respect to the model's own
behavior on the full corpus, hence a high value does not indicate necessarily a
good face validity of the maps with respect to neuroscience knowledge.
The solid line represents the mean across the 13 words and the error bands
represent a 95\% confidence interval based on 1\,000 bootstrap repetitions.
}%
\label{fig:coef-consistency}
\end{minipage}%
\end{figure}
Quantitatively comparing methods on very rare terms is difficult for
lack of ground truth. We therefore conduct meta-analyses on subsampled
corpora, in which some terms are made artificially rare, and use the maps
obtained from the full corpus as a reference.
We choose a set of frequent and well-mapped terms, such as
``language'', for which NeuroQuery and NeuroSynth (trained on a full corpus)
give consistent results. For each of those terms, we construct a series of
corpora in which the word becomes more and more rare: from a full
corpus, we erase randomly the word from many documents until it occurs at most
in $2^{13} = 8912$ articles, then $2^{12} = 4096$, and so on.
For many terms, NeuroQuery only needs a dozen examples to produce maps that are
qualitatively and quantitatively close to the maps it obtains for the full
corpus -- and to NeuroSynth's full-corpus maps. NeuroSynth
typically needs hundreds of examples to obtain similar results, as seen in
\cref{fig:coef-consistency}.
Document frequencies roughly follow a power law \citep{piantadosi2014zipf},
meaning that most words are very rare -- half the terms in our vocabulary occur
in less than 76 articles (see \cref{fig:zipf-law-appendix} in
\cref{document-frequencies}).
Reducing the number of studies required to map well a term (\aka the sample
complexity of the meta-analysis model) therefore greatly widens the vocabulary
that can be studied by meta-analysis.

%

\begin{figure}[!tb]
\begin{minipage}[b]{.4\linewidth}
  \caption{\textbf{Taming query variability} Maps obtained
for a few words related to mental arithmetic. By correctly capturing the fact
that these words are related, NeuroQuery can use its map for easier words like
``calculation'' and ``arithmetic'' to encode terms like ``computation'' and
``addition'' that are difficult for meta-analysis.
  \label{fig:maps-related-to-calculation}%
  \vspace*{-2ex}
  }%
\end{minipage}%
\hfill%
  \iffinal
    \includegraphics[width=.58\textwidth]{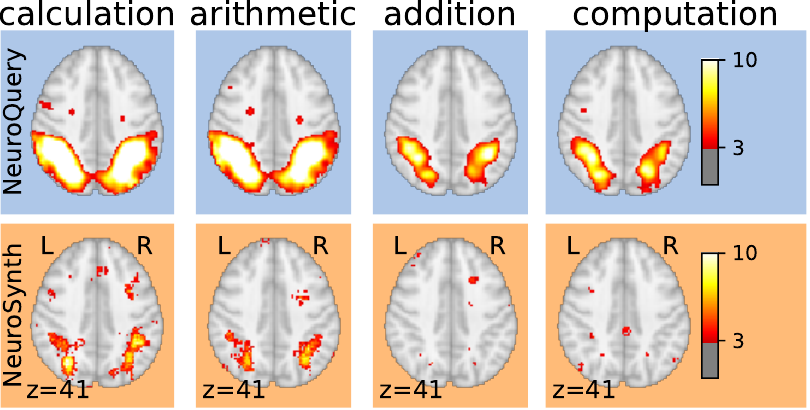}
  \fi%
\end{figure}

\para
Capturing
relations between terms is important because the literature does not use
a perfectly consistent terminology.
The standard solution is to use
expert-built ontologies \citep{poldrack2016brain}, but these tend to have low coverage.
For example, the controlled
vocabularies that we use display relatively small intersections, as can be seen in \cref{table:voc-intersections}. In
addition, ontologies are typically even more incomplete in listing
relations across terms.
Rather than ontologies, 
NeuroQuery relies on distributional semantics and co-occurrence statistics
across the literature to estimate relatedness between terms. 
These continuous semantic links provide robustness to inconsistent
terminology: consistent meta-analytic maps for similar
terms.
For instance, 
``calculation'', ``computation'', ``arithmetic'', and ``addition''
are all related terms that are associated with similar
maps by NeuroQuery. On the contrary, standard automated meta-analysis frameworks
map these terms in isolation, and thus suffer from a lack of statistical 
power and produce empty, or nearly empty, maps for some of these terms
(see \cref{fig:maps-related-to-calculation}).

\para
NeuroQuery improves mapping not only for rare terms that are variants of
concepts widely studied, but also for some concepts rarely studied, such
as ``color'' or ``Huntington'' (\autoref{fig:example-maps}). The main reason is
the semantic smoothing described in \cref{overview-of-the-neuroquery-model}.
Another reason is that working with the full text of publications associates
many more studies to a query: 2779 for ``color'', while NeuroSynth matches only
236 abstracts, and 147 for ``huntington'', a term not known to NeuroSynth.
Full-text matching however requires to give unequal weight to studies, to avoid
giving too much weight to studies weakly related to the query. These weights are
computed by the supervised-learning ridge regression: 
in its dual formulation,
ridge regression is seen as giving weights to training samples \citep[sec
6.1]{bishop2006pattern}.



\subsection{Quantitative evaluation: NeuroQuery is an accurate model of the
literature.}\label{encoding-papers-experiment}

\begin{figure}[!h]
  \hspace{-50pt}
  \begin{minipage}[t]{\textwidth}
  \iffinal
    \includegraphics[width=1.2\textwidth]{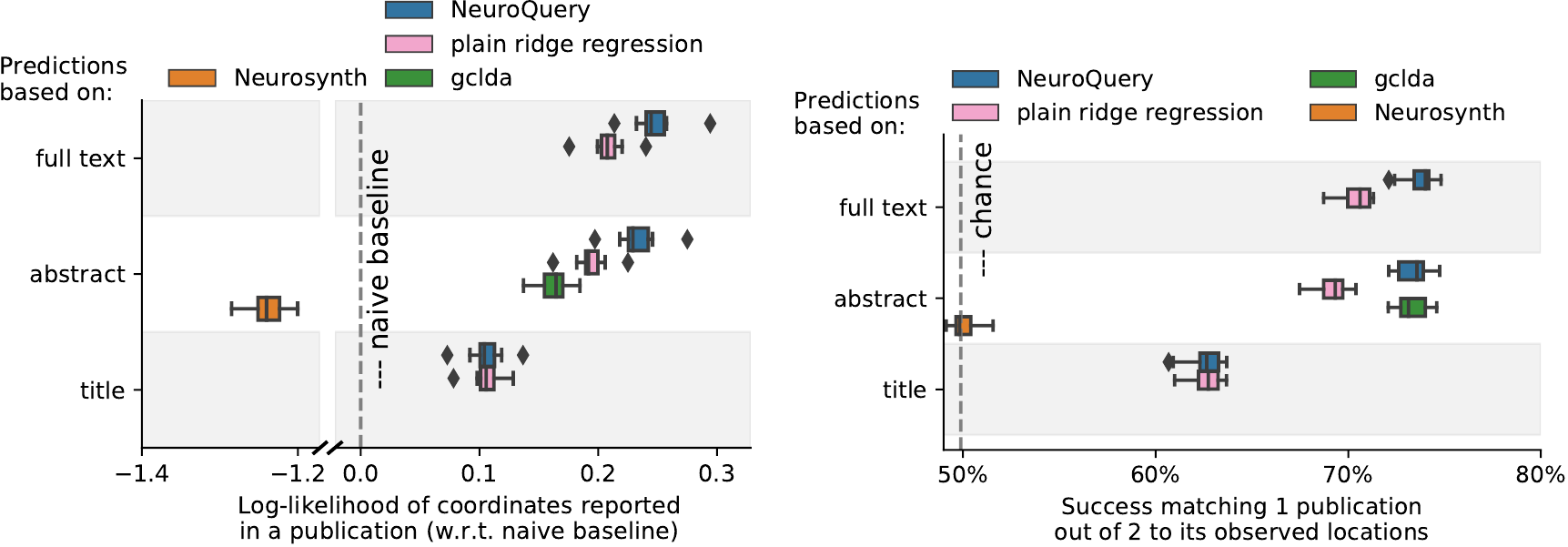}
  \fi%
  \end{minipage}

  \begin{minipage}[t]{\textwidth}
  \caption{\textbf{Explaining coordinates reported in unseen
    studies.}
    -- Left: 
    log-likelihood for coordinates reported  in test articles, relative to
    the log-likelihood of a naive baseline that predicts the average density of the training set. NeuroQuery outperforms
    \ac{GCLDA}, NeuroSynth, and a ridge regression baseline. Note that
    NeuroSynth is not designed to make encoding predictions for full documents,
    which is why it does not perform well on this task. -- Right: how often the predicted map is closer to the true coordinates than to
    the coordinates for another article in the test set
    \citep{mitchell2008predicting}.
    The boxes represent the first, second and third quartiles of scores across 16
    cross-validation folds. Whiskers represent the rest of the distribution,
    except for outliers, defined as points beyond 1.5 times the \acs{IQR} past
    the low and high quartiles, and represented with diamond fliers.
  }
  \label{fig:log-likelihood-box-plot}
  \end{minipage}
\end{figure}

Unlike standard meta-analysis methods, which compute in-sample summary
statistics, NeuroQuery is a predictive model, that can produce brain maps for
out-of-sample neuroimaging studies. This enables us to quantitatively assess its
generalization performance.
Here we check that NeuroQuery captures reliable links from concepts to brain
activity -- associations that generalize to new, unseen neuroimaging studies. We
do this with 16-fold shuffle-split cross-validation.
After fitting a NeuroQuery
model on 90\% of the corpus, for each document in the left-out test set (around
1\,300), we encode it, normalize the predicted brain map to coerce it into a
probability density, and compute the average log-likelihood of the
coordinates reported in the article with respect to this density.
The procedure is then repeated 16 times and
results are presented in \cref{fig:log-likelihood-box-plot}.
We also perform this procedure with NeuroSynth and \ac{GCLDA}. NeuroSynth does
not perform well for this test. Indeed, the NeuroSynth model is designed for
single-phrase meta-analysis, and does not have a mechanism to combine words and
encode a full document. Moreover, it is a tool for in-sample statistical
inference, which is not well suited for out-of sample prediction. \ac{GCLDA}
performs significantly better than chance, but still worse than a simple ridge
regression baseline. This can be explained by the unrealistic modelling of brain
activations as a mixture of a small number of Gaussians, which results in low
spatial resolution, and by the difficulty to perform posterior inference for
\ac{GCLDA}.
Another metric, introduced in \cite{mitchell2008predicting} for encoding models,
tests the ability of the meta-analytic model to match the text of a left-out
study with its brain map. For each article in the test set, we draw randomly
another one and check whether the predicted map is closer to the correct map
(containing peaks at each reported location) or to the random negative example.
More than 72\% of the time, NeuroQuery's output has a higher Pearson correlation
with the correct map than with the negative example (see
\cref{fig:log-likelihood-box-plot} right).

\subsection{NeuroQuery maps reflect well other meta-analytic maps}

The above experiments quantify how well NeuroQuery captures the
information in the literature, by
comparing predictions to reported coordinates.
However, the scores are difficult to interpret, as peak
coordinates reported in the literature
are noisy and incomplete with respect to the full
activation maps.
We also want to quantify the quality of the brain maps generated by
NeuroQuery, extending the visual comparisons of  \cref{fig:example-maps}.
For this purpose, we compare 
NeuroQuery predictions to a few reliable references.

First,
we use a set of diverse and curated \ac{IBMA} maps available publicly
\citep{varoquaux2018atlases}.
This collection contains 19 \ac{IBMA} brain maps, labelled 
with cognitive concepts
such as ``visual words''. For each of these labels, we obtain a prediction from
NeuroQuery and compare it to the corresponding \ac{IBMA} map.
The \ac{IBMA} maps are thresholded. We evaluate whether thresholding the
NeuroQuery predicted maps can recover the above-threshold voxels in
the \ac{IBMA}, quantifying false detections and misses for all thresholds
with the Area Under the \ac{ROC} Curve \citep{fawcett2006introduction}.
NeuroQuery predictions match well the \ac{IBMA} results, with a median \ac{AUC}
of 0.80. Such results cannot be directly obtained with NeuroSynth, as many
labels are missing from NeuroSynth's vocabulary. Manually reformulating the
labels to terms from NeuroSynth's vocabulary gives a median \ac{AUC} of .83 for
NeuroSynth, and also raises the \ac{AUC} to .88 for NeuroQuery (details in
\cref{subsection:comparison-with-brainpedia} and
\cref{fig:brainpedia-scores-with-maps}).

We also perform a similar experiment for anatomical terms, relying on the
Harvard-Oxford structural atlases \citep{desikan2006automated}. Both NeuroSynth
and NeuroQuery produce maps that are close to the atlases' manually segmented
regions, with a median \ac{AUC} of 0.98 for NeuroQuery and 0.95 for NeuroSynth,
for the region labels that are present in NeuroSynth's vocabulary. Details are
provided in \cref{subsection:harvard-oxford-experiment} and
\cref{fig:harvard-oxford}.

For frequent-enough terms, we consider NeuroSynth as a reference. Indeed,
while the goal of NeuroSynth is to reject a voxel-level association test,
and not to predict a activation distribution like NeuroQuery, it would
still be desirable that NeuroQuery predicts few observations where an
association statistic is small.
We threshold NeuroSynth maps by controlling the \ac{FDR} at
1\% and select the 200 maps with the largest number of activations. We compare
NeuroQuery predictions to NeuroSynth activations by computing the \ac{AUC}.
NeuroQuery and NeuroSynth maps for these well-captured terms are very similar,
with a median \ac{AUC} of 0.90. Details are provided in
\cref{subsection:comparison-with-neurosynth} and
\cref{fig:neuroquery-comparison-with-neurosynth}.

\subsection{NeuroQuery is an openly available resource}
NeuroQuery can easily be used online: \url{https://neuroquery.org}.
Users can enter free text in a search box (rather than select a single term from
a list as is the case with existing tools) and discover which terms,
neuroimaging publications, and brain regions are related to their query.
NeuroQuery is also available as an open-source Python package that can be easily
installed on all platforms: \url{https://github.com/neuroquery/neuroquery}.
This will enable advanced users to run extensive meta-analysis with
Neuroquery, integrate it in other
applications, and extend it. The package allows training
new NeuroQuery models as well as downloading and using a pre-trained model.
Finally, all the resources used to build NeuroQuery are freely available
at
\url{https://github.com/neuroquery/neuroquery_data}. This repository contains 
\emph{i)} the data used to train the model:
vocabulary list and document frequencies, word counts (\ac{TFIDF}
features), and peak activation coordinates for our whole
corpus of 13\,459 publications, \emph{ii)} the semantic-smoothing matrix,
that encodes relations across the terminology. The corpus is
significantly richer than NeuroSynth, the largest corpus to date (see
\cref{table:comparison-with-neurosynth-dataset} for a comparison), and
manual quality assurance reveals more accurate extraction of brain
coordinates (\cref{table:coordinate-extraction-errors}).

\section{Discussion and conclusion}

NeuroQuery makes it easy to perform meta-analyses of arbitrary questions on 
the human neuroscience literature: it
uses a full-text description of the question and the studies and it
provides an online query interface with a rich database of studies.
For this, it departs from existing meta-analytic frameworks by
treating meta-analysis as a prediction problem.
It describes neuroscience concepts of interest by continuous combinations
of terms rather than matching publications for exact terms.
As it combines multiple terms and interpolates between available studies,
it extends the scope of meta-analysis in neuroimaging.
In particular,
it can capture information for concepts studied much less frequently than those
that are covered by current automated meta-analytic approaches.

\subsection{Related work}

A variety of prior works have paved the way for NeuroQuery. Brainmap
\citep{laird2005brainmap} was the first systematic database of brain
coordinates. NeuroSynth \citep{yarkoni2011large} pioneered automated
meta-analysis using abstracts from the literature, broadening a lot the
set of terms for which the consistency of reported locations can be
tested. These works perform classic meta-analysis, which considers terms
in isolation, unlike NeuroQuery. Topic models have also been used to find
relationships across terms used in meta-analysis.
\cite{nielsen2004mining} used a non-negative matrix factorization on the
matrix of occurrences of terms for each brain location (voxel): their
model outputs a set of seven spatial networks associated with cognitive topics,
described as weighted combinations of terms.
\cite{poldrack2012discovering} used topic models on the full text of
5\,800 publications to extract from term cooccurrences 130 topics on
mental function and disorders, followed by a classic meta-analysis to map
their neural correlates in the literature. These topic-modeling works
produce a reduced number of cognitive latent factors --or topics-- mapped
to the brain, unlike NeuroQuery which strives to map \emph{individual}
terms and uses
their cooccurences in publications only to infer the semantic links. From
a modeling perspective, the important difference of NeuroQuery is
supervised learning, used as an encoding model
\citep{naselaris2011encoding}. In this sense, the supervised learning
used in NeuroQuery differs from that used in \cite{yarkoni2011large}: the
latter is a decoding model that, given brain locations in a study,
predicts the likelihood of neuroscience terms without using relationships
between terms. Unlike prior approaches, the maps of NeuroQuery are
predictions of its statistical model, as opposed to model parameters.
Finally, other works have modelled co-activations and interactions between brain
locations \citep{kang2011meta,wager2015bayesian,xue2014identifying}. We do not
explore this possibility here, and except for the density estimation NeuroQuery
treats voxels independently.

\subsection{Usage recommendations and limitations}

We have thoroughly validated that NeuroQuery gives quantitatively and
qualitatively good results that summarize well the literature.
Yet, the tool has strengths and weaknesses that should inform its usage.
Brain maps produced by NeuroQuery are predictions, and a specific
prediction may be wrong although the tool
performs well on average. A NeuroQuery
prediction by itself therefore does not support definite
conclusions as it does not come with a statistical test.
Rather, NeuroQuery will be most successfully used to produce hypotheses
and as an exploratory tool, to be confronted with other sources of
evidence.
To prepare a new functional neuroimaging study, NeuroQuery
helps to formulate hypotheses, defining \ac{ROI}s or other
formal priors (for Bayesian analyses).
To interpret results of a neuroimaging experiment, NeuroQuery can readily
use the description of the experiment to assemble maps from the
literature, which can be compared against, or updated using, experimental findings.
As an exploratory tool, extracting patterns from published neuroimaging
findings can help conjecture relationships across mental processes as well as
their neural correlates \citep{yeo2014functional}.
NeuroQuery can also facilitate literature reviews: given a query, it uses its
semantic model to list related studies and their reported activations.
What NeuroQuery does not do is provide conclusive evidence that a brain region is
recruited by a mental process or affected by a pathology.
Compared to traditional meta-analysis tools,
NeuroQuery is particularly beneficial \emph{i)} when the term of
interest is rare, \emph{ii)} when the concept of
interest is best described by a combination of multiple terms, and
\emph{iii)} when a fully automated method is necessary and queries would otherwise need cumbersome
manual curation to be understood by other tools.

\para

Understanding the components of NeuroQuery helps interpreting its results.
We now describe in details potential failures of the tool, and how to
detect them.
NeuroQuery builds predictions by combining brain maps each
associated with a keyword related to the query.
A first step to interpret results is to inspect this list of keywords,
displayed by the online tool. These keywords are selected based on their semantic
relation to the query, and as such will usually be relevant. However, in rare cases, they may build upon undesirable associations. For example, ``agnosia'' is linked to
``visual'', ``fusiform'', ``word'' and ``object'',
because visual agnosia is the type of agnosia most studied in the
literature, even though ``agnosia'' is a much more general concept.
In this specific case, the indirect
association is problematic because ``agnosia'' is not a \emph{selected}
term  that NeuroQuery can map by itself, as it is not well-represented in the source data.
As a result, the NeuroQuery prediction for ``agnosia'' is driven
by indirect associations, and focuses on the visual system, rather than
areas related to, e.g., auditory agnosia. By contrast, ``aphasia'' is an
example of a term that is well mapped, building on maps for
``speech'' and ``language'', terms that are semantically close to
aphasia and well captured in the literature.
%
%

A second consideration is that, in some extreme cases, the semantic smoothing fails to produce meaningful
results. This happens
when a term has no closely related terms that correlate well with brain
activity.
For instance,
``ADHD'' is very similar to ``attention deficit hyperactivity disorder'',
``hyperactivity'', ``inattention'', but none of these terms is selected as a
feature mapped in itself, because their link with brain activity is relatively loose.
Hence, for ``ADHD'', the model builds its
prediction on terms that are distant from the query, and
produces a misleading map
that highlights mostly the
cerebellum\footnote{\url{https://neuroquery.org/query?text=adhd}}.
While this result is not satisfying, the failure is detected by the
NeuroQuery interface and reported with a warning stating that results may not
be reliable.
To a user with general knowledge in psychology, the failure can also be
seen by inspecting the associated terms, as displayed in the user
interface.

A third source of potential
failure stems from NeuroQuery's model of additive combination. This model is not unique to NeuroQuery, and lies
at the heart of functional
neuroimaging, which builds upon the hypothesis of
pure insertion of cognitive processes
\citep{ulrich1999donders,poldrack201013substraction}. An inevitable consequence is that, in some cases, a group of words will not be well mapped by its constituents.
For example, ``visual sentence comprehension'' is decomposed into two
constituents known to Neuroquery: ``visual'' and ``sentence
comprehension''. Unfortunately, the map corresponding to the combination
is then dominated by the primary visual cortex, given that it leads to
very powerful activations in fMRI. Note that ``visual word
comprehension'', a slightly more common subject of interest, is
decomposed into ``visual word'' and ``comprehension'', which leads to a
more plausible map, with strong loadings in the visual word form area.

A careful user can check that each constituent of a query
is associated with a plausible map, and
that they are well combined. The NeuroQuery interface enables to gauge the
quality of the mapping of each individual term by
presenting the corresponding brain map as well as the number of
associated studies.
The final combination can be understood by inspecting the weights of the
combination as well as comparing the final combined map with the maps for
individual terms.
Such an inspection can for instance reveal that, as mentioned above,
``visual'' dominates ``sentence comprehension'' when mapping
``visual sentence comprehension''.

We have attempted to provide a comprehensive overview of the main pitfalls users are likely to encounter when using NeuroQuery, but we hasten to emphasize that all of these pitfalls are infrequent. NeuroQuery produces
reliable maps for the typical queries, as quantified by our experiments.

\subsection{General considerations on meta-analyses}

When using NeuroQuery to foster scientific progress, it is useful to keep
in mind that meta-analyses are not a silver bullet. First,
meta-analyses have little or no ability to correct biases present in the primary literature (e.g., perhaps confirmation bias drives researchers to overreport amygdala activation in emotion studies).
Beyond increased statistical power, one promise of meta-analysis is to
afford a wider perspective on results---in particular, by comparing brain
structures detected across many different conditions. 
However, claims that a structure is selective to a mental condition
need an explicit statistical model of reverse inference
\citep{wager2016pain}. Gathering such evidence is challenging:
selectivity means that
changes at the given brain location specifically \emph{imply} a mental
condition, while brain imaging experiments most often do not manipulate the
brain itself, but rather the experimental conditions it is placed in \citep{poldrack2006can}. In a meta-analysis,
the most important confound for reverse inferences is that some brain
locations are reported for many different conditions.
NeuroQuery accounts for this varying baseline across the brain by fitting
an intercept and reporting only differences from the baseline.
While helpful, this is not a formal statistical test of reverse inference.
For example, the NeuroQuery map for ``interoception'' highlights the insula,
because studies that mention ``interoception'' tend to mention and report
coordinates in the insula. This, of course, does \emph{not} mean that
interoception is the only function of the insula.
Another fundamental challenge of meta-analyses in psychology is the
decomposition of the tasks in mental processes: the descriptions of the
dimensions of the experimental paradigms are likely imperfect and
incomplete. Indeed, even for a task as simple as finger tapping, minor
variations in task design lead to reproducible variations in neural responses
\citep{witt2008functional}.
However, quantitatively describing all aspects of all
tasks and cognitive strategies is presently impossible, as it would
require a universally-accepted, all-encompassing psychological ontology. 
Rather, NeuroQuery grounds meta-analysis in
the full-text descriptions of the studies, which in our view provide the best available
proxy for such an idealized ontology.


\subsection{Conclusion}

NeuroQuery stems from a desire to
compile results across studies and laboratories, an essential endeavor for the
progress of human brain mapping \citep{yarkoni2010cognitive}. Mental
processes are difficult to isolate and findings of individual studies 
may not generalize. Thus, tools are needed to denoise and summarize
knowledge accumulated across a large number of studies.
Such tools must be usable in practice and match the needs of researchers who
exploit them to study human brain function and disorders.
NeuroSynth took a huge step in this direction by enabling anyone to perform, in
a few seconds, a fully automated meta-analysis across thousands of studies, for
an important number of isolated terms.
Still, users are faced with the difficult task of mapping their question to a
single term from the NeuroSynth vocabulary, which cannot always be done in a
meaningful way. If the selected term is not popular enough,
the resulting map also risks being unusable for lack of statistical power.
NeuroQuery provides statistical maps for arbitrary queries -- from
seldom-studied terms to free-text descriptions of experimental protocols. Thus, it
enables applying
fully-automated and quantitative meta-analysis in situations where
only semi-manual and subjective solutions were available.
It therefore
brings an important advancement towards grounding neuroscience on 
quantitative knowledge representations.

\paragraph{Acknowledgements}: JD acknowledges funding from Digiteo under
project Metacog (2016-1270D). TY acknowledges funding from NIH under
grant number R01MH096906. BT received funding from the European Union's Horizon 2020
Research and Innovation Programme under Grant Agreement No. 785907 (HBP SGA2)
and No 826421 (VirtualbrainCloud). FS acknowledges funding from ANR
via grant ANR-16- CE23-0007-01 (``DICOS''). GV was partially funded by
the Canada First Research Excellence Fund, awarded to McGill University
for the Healthy Brains for Healthy Lives initiative.
We also thank the reviewers, including Tor D. Wager, for their suggestions
that improved the manuscript.

\bibliography{cognitive_science_paper.bib}

\clearpage
\appendix
\part{Supplementary material} 
\parttoc 

\section{Building the NeuroQuery training data}\label{section:building-neuroquery-data}

\subsection{A new dataset}
The dataset collected by NeuroSynth \citep{yarkoni2011large} is openly
available\footnote{\url{https://github.com/neurosynth/neurosynth-data}}.
In July, 2019, NeuroSynth contains 448\,255 unique locations for
14\,371 studies. It also contains the term frequencies for 3228 terms (1335 are
actually used in the NeuroSynth online
tool\footnote{\url{http://neurosynth.org}}), based on the abstracts of the
studies.
However, it only contains term frequencies for the abstracts, and not the
articles themselves. This results in a shallow description of the studies, based
on a very short text (around 20 times smaller than the full article).
As a result, many important terms are very rare: they seldom occur in
abstracts, and can be associated with very few studies.
For example, in our corpus of 13\,459 studies, ``huntington disease''
occurs in 32 abstracts, and ``prosopagnosia'' in 25.
For such terms, meta-analysis lacks statistical power.
When the full text is available, many more term occurrences -- associations
between a term and a study -- are observed
(\cref{fig:scatter-doc-frequencies-neuroquery-vs-neurosynth}). This means that
more information is available, terms are better described by their set of
associated studies, and meta-analyses have more statistical power.
Moreover, as publications cannot always be redistributed for
copyright reasons, NeuroSynth (and any dataset of this nature) can only provide
term frequencies for a fixed vocabulary, and not the text they were extracted
from.
We therefore decided to collect a new corpus of neuroimaging studies, which
contains the full text. We also created a new peak activation coordinate
extraction system, which achieved a higher precision and recall than
NeuroSynth's on a small sample of manually annotated studies.

\subsection{Journal articles in a uniform and validated format}

We downloaded around 149\,000 full-text journal articles related to neuroimaging
from the PubMed Central\footnote{\url{https://www.ncbi.nlm.nih.gov/pmc/},
\url{https://www.ncbi.nlm.nih.gov/books/NBK25501/}} \citep{sayers2009utilities}
and Elsevier\footnote{\url{https://dev.elsevier.com/api_docs.html}} APIs.
We focus on these sources of data because they provide many articles in a
structured format. It should be noted that this could result in a selection
bias, as some scientific journals -- mostly paid journals -- are not available
through these channels.
The articles are selected by querying the ESearch Entrez utility
\citep{sayers2009utilities} either for specific neuroimaging journals or with
query strings such as ``fMRI''. The resulting studies are mostly based on
\ac{fMRI} experiments, but the dataset also contains \ac{PET} or structural
\ac{MRI} studies. It contains studies about diverse types of populations:
healthy adults, patients, elderly, children.

We use \ac{XSLT} to convert all articles to the \ac{JATS} Archiving and
Interchange XML language\footnote{\url{https://jats.nlm.nih.gov/archiving/}} and
validate the result using the W3C XML Schema (XSD) schemas provided on the
\ac{JATS} website.
From the resulting XML documents, it is straightforward to extract the title,
keywords, abstract, and the relevant parts of the article body, discarding the
parts which would add noise to our data (such as the acknowledgements or
references).

\subsection{Coordinate extraction}\label{subsection:coordinate-extraction}
We extract tables from the downloaded articles and convert them to the XHTML 1.1
table model (the \ac{JATS} also allows using the OASIS CALS table model). We use
stylesheets provided by docbook\footnote{\url{https://docbook.org/tools/}} to
convert from CALS to XHTML.
Cells in tables can span several rows and columns. When extracting a table, we
normalize it by splitting cells that span several rows or columns and
duplicating these cells' content; the normalized table thus has the shape of a
matrix.
Finally, all unicode characters that can be used to represent ``+'' or ``-''
signs (such as \verb|&#x2212;| ``MINUS SIGN'') are mapped to their ASCII
equivalents, ``+'' (\verb|&#x2b;| ``PLUS SIGN'') or ``-'' (\verb|&#x2d;|
``HYPHEN MINUS'').
Once tables are isolated, in XHTML format, and their rows and columns are
well aligned, the last step is to find and extract peak activation
coordinates.
Heuristics find columns containing either single coordinates or triplets of
coordinates based on their header and the cells' content.
A heuristic detects when the coordinates extracted from a table are probably not
stereotactic peak activation coordinates, either because many of them lie
outside a standard brain mask, or because the group of coordinates as a whole
fits a normal distribution too well. In such cases the whole table is discarded.
Out of the 149\,000 downloaded and formatted articles, 13\,459 contain
coordinates that could be extracted by this process, resulting in a total of
418\,772 locations.

All the extracted coordinates are treated as coordinates in the \ac{MNI} space,
even though some articles still refer to the Talairach space. The precision of
extracted coordinates could be improved by detecting which reference is used and
transforming Talairach coordinates to \ac{MNI} coordinates. However, differences
between the two coordinate systems are at most of the order of \SI{1}{cm}, and
much smaller in most of the brain. This is comparable to the size of the
Gaussian kernel used to smooth images. Moreover, the alignment of brain images
does not only depend on the used template but also on the registration method,
and there is no perfect transformation from Talairach to \ac{MNI} space
\citep{lancaster2007bias}. Therefore, treating all coordinates uniformly is
acceptable as a first approximation, but better handling of Talairach
coordinates is a clear direction for improving the NeuroQuery dataset.

\paragraph{Coordinate extraction evaluation.}\label{paragraph:coordinate-extraction-evaluation}
To evaluate the coordinate extraction process, we focused on articles that are
present in both NeuroSynth's dataset and NeuroQuery's, and for which the two
coordinate extraction systems disagree.
Out of 8\,692 articles in the intersection of both corpora, the extracted
coordinates differ (for at least one coordinate) in 1\,961 (i.e.\; in 23\% of
articles).
We selected the first 40 articles (sorted by PubMed ID) and manually evaluated
the extracted coordinates.
As shown in \cref{table:coordinate-extraction-errors}, our method extracted false
coordinates from fewer articles: 3 / 40 articles have at least one false
location in our dataset, against 20 for NeuroSynth. While these numbers may seem
high, note that errors are far less likely to occur in articles for which both
methods extract exactly the same locations.
\begin{table}
  \begin{tabular}{lll}
    & False positives & False negatives\\
    NeuroSynth & 20 & 28\\
    NeuroQuery & 3 & 8
  \end{tabular}
  \caption{Number of extracted coordinate sets that contain at least one error
    of each type, out of 40 manually annotated articles. The articles are chosen
    from those on which NeuroSynth and NeuroQuery disagree -- the ones most
    likely to contain errors.}\label{table:coordinate-extraction-errors}
\end{table}

\subsection{Density maps}
For each article, the coordinates from all tables are pooled, resulting in a
set of peak activation coordinates. We then use Gaussian \ac{KDE}
\citep{silverman1986density,scott2015multivariate}
to estimate the density of these activations over the brain. The chosen bandwidth of
the Gaussian kernel yields a \ac{FWHM} close to 9mm, which is in the range of
smoothing kernels that are typically used for \ac{fMRI} meta-analysis
\citep{wager2007meta,wager2004neuroimaging,turkeltaub2002meta}.
For comparison, NeuroSynth uses a hard ball of 10mm radius.

One benefit of focusing on the density of peak coordinates (which is
$\ell_1$-normalized) is that it does not depend on the number of contrasts
presented in an article, nor on other analytic choices that cause the number of
reported coordinates to vary widely, ranging from less than a dozen to several
hundreds.

\subsection{Vocabulary and \ac{TFIDF} features}

We represent the text of our articles by \ac{TFIDF} features
\citep{salton1988term}. These simple representations are popular in document
retrieval and text classification because they are very efficient for many
applications. They contain the (reweighted) frequencies of many terms in the
text, discarding the order in which words appear.
An important choice when building \ac{TFIDF} vectors is the vocabulary: the
words or expressions whose frequency are measured. It is common to use all words
encountered in the training corpus, possibly discarding those that are too
frequent or too rare.
The vocabulary is often enriched with ``n-grams'', or collocations: groups
of words that often appear in the same sequence, such as ``European Union'' or
``default mode network''. These collocations are assigned a dimension of the
\ac{TFIDF} representations and counted as if they were a single token. There are
several strategies to discover such collocations in a training corpus
\citep{mikolov2013distributed, bouma2009normalized}.

We do not extract the vocabulary and collocations from the training corpus, but
instead rely on existing, manually-curated vocabularies and ontologies of
neuroscience. This ensures that we only consider terms that are relevant to
brain function, anatomy or disorders, and that we only use meaningful
collocations. Moreover, it helps to reduce the dimensionality of the \ac{TFIDF}
representations.
Our vocabulary comprises five important lexicons of neuroscience, based on
community efforts: the subset of \acf{MeSH} (\url{https://www.ncbi.nlm.nih.gov/mesh})
dedicated to neuroscience and psychology, detailed in \cref{parts-of-mesh-used}
(\ac{MeSH} are the terms used by PubMed
to index articles), Cognitive Atlas (\url{http://www.cognitiveatlas.org/}), NeuroNames
(\url{http://braininfo.rprc.washington.edu/NeuroNames.xml}) and NIF
(\url{https://neuinfo.org/}). We also include all the terms and bigrams used by
NeuroSynth (\url{http://neurosynth.org}). We discard all the terms and expressions
that occur in less than
5 / 10\,000 articles. The resulting vocabulary contains 7\,547 terms and
expressions related to neuroscience.

\subsection{Summary of collected data}

The data collection described in this section provides us with important
resources:
\emph{i)}
Over 149K full-text journal articles related to neuroscience -- 13.5K of which
contain peak activation coordinates -- all translated into the same structured
format and validated.
\emph{ii)}
Over 418K peak activation coordinates for more than 13.5K articles.
\emph{iii)}
A vocabulary of 7547 terms related to neuroscience, each occurring in at least 6
articles from which we extracted coordinates.
This dataset is the largest of its kind. In what follows we focus on the set of
13.5K articles from which we extracted peak locations.

Some quantitative aspects of the NeuroQuery and NeuroSynth datasets are
summarized in \cref{table:comparison-with-neurosynth-dataset}.

\begin{table}[h!]
\begin{tabular}{lrr}
 & NeuroSynth & NeuroQuery\\
\hline
Dataset size &  & \\
\hline
articles & 14\,371 & 13\,459\\
terms & 3\,228 (1\,335 online) & 7\,547\\
journals & 60 & 458\\
raw text length (words) & $\approx$ 4 M & $\approx$ 75 M\\
unique term occurrences & 1\,063\,670 & 5\,855\,483\\
unique term occurrences in voc intersection & 677\,345 & 3\,089\,040\\
coordinates & 448\,255 & 418\,772\\
\hline
Coordinate extraction errors on conflicting articles &  & \\
\hline
articles with false positives / 40 & 20 & 3\\
articles with false negatives / 40 & 28 & 8\\
\end{tabular}
  \caption{Comparison with NeuroSynth. ``voc intersection'' is the set of terms
    present in both NeuroSynth's and NeuroQuery's vocabularies. The
    ``conflicting articles'' are papers present in both datasets, for which the
    coordinate extraction tools disagree, 40 of which were manually annotated.}\label{table:comparison-with-neurosynth-dataset}
\end{table}

\paragraph{Text.}
In terms of raw amount of text, this corpus is 20 times larger
than NeuroSynth's. Combined with our vocabulary, it yields over 5.5M
occurrences of a unique term in an article. This is over 5 times more than the
word occurrence counts distributed by
NeuroSynth\footnote{\url{https://github.com/neurosynth/neurosynth-data}}. When
considering only terms in NeuroSynth's vocabulary, the corpus still contains
over 3M term-study associations, 4.6 times more than NeuroSynth.
Using this larger corpus results in denser representations,
higher statistical power, and coverage of a wider vocabulary.
There is an important overlap between the selected studies: 8\,692 studies are
present in both datasets -- the Intersection Over Union is 0.45.

\paragraph{Coordinates.}
The set of extracted coordinates is almost the size of NeuroSynth's (which is
7\% larger with 448\,255 coordinates after removing duplicates), and is less
noisy. To compare coordinate extractions, we manually annotated a small set of
articles for which NeuroSynth's coordinates differ from NeuroQuery's. Compared
with NeuroSynth, NeuroQuery's extraction method reduced the number of articles
with incorrect coordinates (false positives) by a factor of 7, and the number of
articles with missing coordinates (false negatives) by a factor of 3
(\cref{table:coordinate-extraction-errors}). Less noisy brain activation data is
useful for training encoding models.
\paragraph{Sharing data.}
We do not have the right to share the full text of the articles, but the
vocabulary, extracted coordinates, and term occurrence counts for the whole
corpus are freely available
online\footnote{\url{https://github.com/neuroquery/neuroquery_data/training_data}}.

\section{Methodological details}\label{methodological-details}

\subsection{Notation}
We denote scalars, vectors and matrices with lower-case, bold lower-case, and
bold-upper case letters respectively: $x$, $\x$, $\X$.
We denote the elements of $\X$ by $x_{i,j}$, its rows by $\x_i$, and its columns
by $\x_{*,i}$.
We denote $p$ the number of voxels in the brain, $v$ the size of the vocabulary,
and $n$ the number of studies in the dataset.
We use indices $i$, $j$, $k$ to indicate indexing samples (studies), features
(terms), and outputs (voxels) respectively.
We use a hat to denote estimated values, \eg $\hat{\B}$.
$\langle \x, \y \rangle$ is the vector scalar product.

\subsection{\acs{TFIDF} feature extraction}
We represent a document by its \ac{TFIDF} features \citep{salton1988term}, which are
reweighted \acl{BOW} features.
A \ac{TFIDF} representation is a vector in which each entry corresponds to the
(reweighted) frequency of occurrence of a particular term.
The \emph{term frequency}, $\tf$, of a word in a document is the number of times
the word occurs, divided by the total number of words in the document.
The \emph{document frequency}, $\df$, of a word in a corpus is the proportion of
documents in which it appears.
The \emph{inverse document frequency}, $\idf$, is defined as:
\begin{equation}
  \idf(w) = - \log(\df) + 1 = - \log \frac{|\,\{\,i \;| \; w \text{ occurs in document } \; i\,\}\,|}
  {n}  + 1 \; ,
\end{equation}
where $n$ is the number of documents in the corpus and $|\cdot|$ is the cardinality.
Term frequencies are reweighted by their $\idf$, so that frequent words, which
occur in many documents (such as ``results'' or ``brain''), are given less
importance. Indeed, such words are usually not very informative.

Our \ac{TFIDF} representation for a study is the uniform average of the
normalized \ac{TFIDF} vectors for its title, abstract, full text, and keywords.
Therefore, all parts of the article are taken into account, but a word that
occurs in the title is more important than a word the article body (since the
title is shorter).

\ac{TFIDF} features exploit a fixed vocabulary -- each dimension
is associated with a particular word.
The vocabulary we consider comprises 7\,547 terms or phrases related to
neuroscience that occur in at least 0.05\% of publications. These terms are
extracted from manually curated sources shown in \cref{table:voc-intersections}
and \cref{table:atlases-used}.

\subsection{Reweighted ridge matrix and feature (vocabulary) selection}\label{reweighted-ridge-regression}

Here we give some details about the feature selection and adaptive ridge
regularization.
After extracting \ac{TFIDF} features and computing density estimation maps, we
fit a linear model by regressing the activity of each voxel on the \ac{TFIDF}
descriptors (\cref{overview-of-the-neuroquery-model}). We denote $p$ the number
of voxels, $v$ the size of the vocabulary, and $n$ the number of documents in
the corpus. We construct a design matrix $ X \in \mathbb{R}^{n \times v}$
containing the \ac{TFIDF} features of each study, and the dependent variables
$\Y \in \mathbb{R}^{n \times p}$ representing the activation density at
each voxel for each study. The linear model thus writes:

\begin{equation}
  \Y = \X \, \B^* + \bm{E},
\end{equation}
where $\bm{E}$ is Gaussian noise and $\B^* \in \mathbb{R}^{v \times
p}$ are the unknown model coefficients. We use ridge regression (least-squares
regression with a penalty on the $\ell_2$ norm of the model coefficients).
Some words are much more informative than others, or have a much stronger
correlation with brain activity. For example, ``auditory'' is well correlated
with activations in the auditory areas, whereas ``attention'' has a lower
signal-to-noise ratio, as it is polysemic and, even when used as a
psychological concept, has a weaker link to reported neural activations.
Therefore it is beneficial to adapt the amount of regularization for each word,
to strongly penalize (or even discard) the most noisy features.

Many existing methods for feature selection are not adapted to our case,
because:
\emph{i)} the design matrix $\bm{X}$ is very sparse, and more importantly
\emph{ii)} we want to select the same features for $\approx$ 28\,000 outputs (each
voxel in the brain is a dependent variable).
We therefore introduce a new reweighted ridge regression and feature selection
procedure.

Our approach is based on the observation that when fitting a ridge regression
with a uniform regularization, the most informative words are associated
with large coefficients for many voxels.
We start by fitting a ridge regression with uniform regularization. We
obtain one statistical map of the brain for every feature (every term in the
vocabulary). The maps are rescaled to reduce the importance of
coefficients with a high variance.
We then compute the squared $\ell_2$ norms of these brain maps across voxels.
These norms are a good proxy for the importance of each feature. Terms
associated with large norms explain well the activity of many voxels and tend to
be helpful features. We rely on these brain map norms to determine
which features are selected and what regularization is applied.
The feature selection and adaptive regularization are described in detail in the
rest of this section.

\subsubsection{Z scores for ridge regression coefficients}\label{subsection:ridge-coef-variance}

Our design matrix $\bm{X} \in \mathbb{R}^{n \times v}$ holds \ac{TFIDF} features
for $v$ terms in $n$ studies. There are $p$ dependent variables, one for each
voxel in the brain, which form $\bm{Y} \in \mathbb{R}^{n \times p}$.
The first ridge regression fit yields coefficients $\hat{\bm{B}}^{(0)} \in
\mathbb{R}^{v \times p}$:
\begin{equation}\label{eq:neuroquery-first-ridge-fit}
  \hat{\bm{B}}^{(0)} =
\argmin_{\bm{B} \in \mathbb{R}^{v \times p}} ||\bm{Y} - \bm{X} \, \bm{B}||_{\F}^2 +
\lambda \, ||\bm{B}||_{\F}^2,
\end{equation} where $\lambda \in \mathbb{R}_{> 0}$ is a hyperparameter set with
\acf{GCV} \citep{rifkin2007notes}.
We then compute an estimate of the variance of these coefficients. The approach
is similar to the one presented in \citet{Gaonkar2012} for the case of
\acs{SVM}s.
A simple estimator can be obtained by noting that the coefficients of a ridge
regression are a linear function of the dependent variables.
Indeed, solving \cref{eq:neuroquery-first-ridge-fit} yields:
\begin{equation}
  \hat{\bm{B}}^{(0)} = (\bm{X}^T\bm{X} + \lambda \bm{I})^{-1}\bm{X}^T \bm{Y} \; .
\end{equation}
Defining
\begin{equation}
  \bm{M} = (\bm{X}^T\bm{X} + \lambda \bm{I})^{-1}\bm{X}^T
  \in \mathbb{R}^{v \times n} \; ,
\end{equation}
for a voxel $k \in \{1, \dots, p\}$, and a feature $j \in \{1, \dots v\}$,
\begin{equation}
  \hat{b}^{(0)}_{j, k} = \langle  \m_j \,, \, \y_{*,k} \rangle \; ,
\end{equation}
where $\m_j \in \mathbb{R}^{n}$ is the $i^{\text{th}}$ row of $\M$ and
\(\y_{*,k} \in \mathbb{R}^n\) is the $k^{\text{th}}$ column of $\Y$.
The activations of voxel $k$ across studies are considered to be \ac{i.i.d}, so
\begin{equation}
  \Var(\y_{*,k}) = \Var(\y_{1,k}) \; \bm{I}_n \triangleq s_k^2 \; \bm{I}_n \; .
\end{equation}
An estimate of this variance can be obtained from the residuals:
\begin{equation}
  \hat{s}_k^2 \triangleq \frac{1}{n} \, \sum_{i=1}^n \, (\hat{y}^{(0)}_{i, k} - y_{i, k})^2 =
\frac{1}{n} \, \sum_{i=1}^n \, ((\X \hat{\B}^{(0)})_{i, k} - y_{i, k})^2 \; .
\end{equation}
A simple estimate of the coefficients' variance is then:
\begin{equation}\label{eq:neuroquery-estimate-of-coef-variance}
\hat{\sigma}_{j,k}^2 \triangleq
  \widehat{\Var}(\hat{b}^{(0)}_{j, k}) =
  \hat{s}_k^2 \; \langle \m_j\,,\,\m_j \rangle \; = \hat{s}_k^2 \; \sum_{i=1}^n m_{j,i}^2
\end{equation}
We can thus estimate the standard deviation of each entry of
$\hat{\bm{B}}^{(0)}$. We obtain a brain map of Z scores for each term in the
vocabulary: for term $j \in \{1, \dots, v\}$ and voxel $k \in \{1 \dots p\}$,
\begin{equation}\label{eq:neuroquery-zi-def}
  \hat{z}_{j,k} \triangleq \frac{\hat{b}^{(0)}_{j,k}}{\hat{\sigma}_{j,k}} \; .
\end{equation}
We denote $\hat{\bm{\sigma}_j} =  (\hat{\sigma}_{j,1}, \, \dots \, , \hat{\sigma}_{j, p}) \in
\mathbb{R}^p$; and the Z-map for term $j$: $\hat{\bm{z}}_{j} =
(\hat{z}_{j,1}, \, \dots \, , \hat{z}_{j, p}) \in \mathbb{R}^p$ .

\subsubsection{Reweighted ridge matrix}

Once we have a Z-map for each term, we summarize these maps by computing their squared
Euclidean norm. In practice, we smooth
the Z scores: $\hat{z}_{j,k}$ in \cref{eq:neuroquery-zi-def} is replaced by
\begin{equation}\label{eq:neuroquery-smoothing-coef-variance-estimates}
   \hat{\zeta}_{j,k} = \frac{\hat{b}^{(0)}_{j,k}}{\hat{\sigma}_{j,k} + \delta} \; ,
\end{equation}
where $\delta$ is a constant offset.
The offset $\delta$ allows us to interpolate between basing the regularization
on the Z scores, or on the raw coefficients, \ie the $\beta$-maps.
We obtain better results with a large value for $\delta$, such as the mean
variance of all the regression coefficients. This prevents selecting terms only
because they have a very small estimated variance in some voxels.
Note that this offset $\delta$ is only used to compute the regularization, and
not to compute the rescaled predictions produced by NeuroQuery as in
\cref{eq:neuroquery-rescaled-predictions}.

We denote $\hat{\bm{\zeta}}_j = (\hat{\zeta}_{j,1}, \dots , \hat{\zeta}_{j,p}) \in
\mathbb{R}^p, \; \forall j \in \{1, \dots, v\}$.
Next, we compute the mean $\mu$ and standard deviation $e$ of
$\{\,||\hat{\bm{\zeta}}_j||_2^2, \, j = 1 \dots v\,\}$,
and set an arbitrary cutoff
\begin{equation}
  \label{eq:definition-of-cutoff-feat-selection}
  c = \mu + 2 \, e \; .
\end{equation}
All features $j$ such that $||\hat{\bm{\zeta}}_j||_2^2 \leq c + \epsilon $,
where $\epsilon$ is a small margin to avoid division by zero in
\cref{eq:neuroquery-wi-def}, are discarded. In practice we set $\epsilon$ to
$0.001$. The value of $\epsilon$ is not important, because features that are not
discarded but have their $\bm{\zeta}$ norm close to $c$ get very heavily
penalized in \cref{eq:neuroquery-wi-def} and have coefficients very close to
$0$.

We denote $u < v$ the number of features that remain in the selected vocabulary.
We denote $\phi : \{1 \dots u\} \rightarrow \{1 \dots v\}$ the strictly increasing
mapping that reindexes the features by keeping only the $u$ selected terms:
$\phi(\{1 \dots u\})$ is the set of selected features.
We denote
$\bm{P} \in \mathbb{R}^{u \times v}$
the corresponding projection matrix:
\begin{equation}
  \label{eq:projection-matrix-associated-with-feature-selection}
\bm{p}^T_{*,j} = \bm{e}_{\phi(j)}\, , \; \forall j \in \{1 \dots u\} \; ,
\end{equation}
where
$\{\bm{e}_j, \, j = 1 \dots v\}$ is the natural basis of $\mathbb{R}^v$.
The regularization for the selected features is then set to
\begin{equation}\label{eq:neuroquery-wi-def}
w_j = \frac{1}{||\hat{\bm{\zeta}}_{\phi(j)}||_2^2 - c} \;.
\end{equation}
Finally, we define the diagonal matrix $\bm{W} \in \mathbb{R}^{u \times u}$ such
that the $j^{\text{th}}$ element of its diagonal is $w_j$ and fit a new set of
coefficients $\hat{\bm{B}} \in \mathbb{R}^{u \times p}$ with this new ridge
matrix.

\subsubsection{Fitting the reweighted ridge regression}
The reweighted ridge regression problem writes:
\begin{equation}
  \hat{\bm{B}} = \argmin_{\bm{B} \in \mathbb{R}^{u \times p}}
  ||\bm{Y} - \bm{X}\,\bm{P}^T\,\bm{B}||_{\F}^2 + \gamma \Tr(\bm{B}^T\, \bm{W} \,\bm{B}) \; ,
\end{equation}
Where $\gamma \in \mathbb{R}_{> 0}$ is a new hyperparameter, that is again set
by \acf{GCV}.
With a change of variables this becomes equivalent to solving the usual ridge
regression problem:
\begin{equation}
  \hat{\bm{\Gamma}} =
  \argmin_{\bm{\Gamma}}||\bm{Y} - \tilde{\bm{X}} \, \bm{\Gamma}||_{\F}^2
  + \gamma \, ||\bm{\Gamma}||_{\F}^2 \; ,
\end{equation}
where
$\tilde{\bm{X}} = \bm{X} \, \bm{P}^T \, \bm{W}^{-\frac{1}{2}}$
and we recover $\hat{\bm{B}}$ as
$\hat{\bm{B}} = \bm{W}^{-\frac{1}{2}} \, \hat{\bm{\Gamma}}$ .

The variance of the parameters $\hat{\bm{B}}$ can be estimated as in
\cref{eq:neuroquery-estimate-of-coef-variance} -- without applying the smoothing
of \cref{eq:neuroquery-smoothing-coef-variance-estimates}.
NeuroQuery can thus report rescaled predictions
\begin{equation}\label{eq:neuroquery-rescaled-predictions}
  \hat{\bm{z}} = \frac{\bm{x}^T\hat{\bm{B}}}{\left(\widehat{\Var}(\bm{x}^T\hat{\bm{B}})\right)^{\frac{1}{2}}}
\end{equation}
One benefit of this rescaling is to provide the user a natural value to
threshold the maps. As visible on figures
\ref{fig:example-maps}, \ref{fig:coef-consistency}, and
\ref{fig:maps-related-to-calculation}, thresholding \eg at
$\hat{\bm{z}} \approx 3$ selects regions typical of the query, that can be used for
instance in a region of interest analysis.

\subsubsection{Summary of the regression with adaptive regularization}
The whole procedure for feature selection and adaptive regularization is
summarized in \cref{algo:reweighted-ridge-regression}.

\begin{algorithm} 
  \SetAlgoLined
  \SetKwInOut{Input}{input}
  \Input{\ac{TFIDF} features $\bm{X}$, brain activation densities $\bm{Y}$,
    regularization hyperparameter grid $\Lambda$, variance smoothing parameter $\delta$}

  use \ac{GCV} to compute the best hyperparameter $\lambda \in \Lambda$
  and $\hat{\bm{B}}^{(0)} = \argmin_{\bm{B}} ||\bm{Y} - \bm{X}\bm{B}||_{\F}^2 + \lambda ||\bm{B}||_{\F}^2$\;

  compute variance estimates $\hat{\bm{\sigma}}_j^2$ as in \cref{eq:neuroquery-estimate-of-coef-variance}\;

  $\hat{\bm{\zeta}}_j \leftarrow \frac{\hat{\bm{b}}_j^{(0)}}{\hat{\bm{\sigma}}_j +
    \delta} \; \forall j \in \{1 \dots v \} $\;

  compute $c$ according to \cref{eq:definition-of-cutoff-feat-selection} \;

  define $\phi$ the reindexing that selects features $j$ such that
  $||\hat{\bm{\zeta}}_j||_2^2 > c + \epsilon $ \;

  define $\bm{P} \in \mathbb{R}^{u \times v}$ the projection matrix for $\phi$
  as in \cref{eq:projection-matrix-associated-with-feature-selection} \;

  $w_j \leftarrow \frac{1}{||\hat{\bm{\zeta}}_{\phi(j)}||_2^2 - c} \; \forall j \in
  \{1 \dots u \} $\;

  $\bm{W} \leftarrow \diag(w_j, \, j=1 \dots u)$\;

  use \ac{GCV} to compute the best hyperparameter $\gamma \in \Lambda$
  and
  $\hat{\bm{B}} = \argmin_{\bm{B}}
  ||\bm{Y} - \bm{X}\bm{P}^T\bm{B}||_{\F}^2 + \gamma \Tr(\bm{B}^T \bm{W} \bm{B}) $ \;
  \Return{
    $\hat{\bm{B}}$,
    $\widehat{\Var}(\hat{\bm{B}})$,
    $\gamma$,
    $\bm{P}$,
    $\bm{W}$
  }
  \caption{Reweighted Ridge Regression}\label{algo:reweighted-ridge-regression}
\end{algorithm}

In practice, the feature selection keeps $u \approx 200$ features.
It has a very low computational cost compared to other feature selection
schemes. The computational cost is that of fitting two ridge regressions (and
the second one is fitted with a much smaller number of features).
Moreover, the feature selection also reduces computation at prediction time,
which is useful because we deploy an online tool based on the NeuroQuery
model\footnote{\url{https://neuroquery.saclay.inria.fr}}.

\subsection{Smoothing: regularization at test time}\label{subsection:smoothing-details}

In order to smooth the sparse input features, we exploit the covariance
of our training corpus.
We rely on \acf{NMF} \citep{lee1999learning}. We use a \ac{NMF} of $\bm{X} \in
\mathbb{R}^{n \times v}$ to compute a low-rank approximation of the covariance
$\bm{X}^T\,\bm{X} \in \mathbb{R}^{v \times v}$. Thus, we obtain a denoised term
co-occurrence matrix, which measures the strength of association between pairs
of terms.
We start by computing an approximate factorization of the corpus \ac{TFIDF} matrix $\bm{X}$:
\begin{equation}\label{eq:nmf-problem}
  \bm{U}, \bm{V} = \argmin_{\substack{\bm{U} \in \mathbb{R}_{\geq 0}^{n \times d} \\ \bm{V} \in \mathbb{R}_{\geq 0}^{d \times v}}}
  ||\bm{X} - \bm{U} \, \bm{V}||_{\F}^2
  + \lambda(||\bm{U}||_{\F}^2 + ||\bm{V}||_{\F}^2)
  + \gamma (||\bm{U}||_{1,1} + ||\bm{V}||_{1,1}) \, ,
\end{equation}
where $d < v$ is a hyperparameter and $||\, \cdot \, ||_{1,1}$ designates the sum
of absolute values of all entries of a matrix.
Computing this factorization amounts to describing each document in the corpus
as a linear mixture of $d$ latent factors, or \emph{topics}. In \acl{NLP},
similar decomposition methods are referred to as \emph{topic modelling}
\citep{deerwester1990indexing,blei2003latent}.

The latent factors, or topics, are the rows of $\bm{V} \in \mathbb{R}^{d \times
  v}$: each topic is characterized by a vector of positive weights over the
terms in the vocabulary. $\bm{U} \in \mathbb{R}^{n \times d}$ contains the
weight that each document gives to each topic.
For each term in the vocabulary, the corresponding column of $\bm{V}$ is a a
$d$-dimensional \emph{embedding} in the low-dimensional, latent space: this
embedding contains the strength of association of the term with each topic.
These embeddings capture semantic relationships: related terms tend to be
associated with embeddings that have large inner products.

The hyperparameters \(d = 300\), \(\lambda = 0.1\) and \(\gamma = 0.01\) are set
by evaluating the reconstruction error, sparsity of the similarity matrix, and
extracted topics (rows of $\bm{V}$) on an unlabelled (separate) corpus. We find
that the NeuroQuery model as a whole is not very sensitive to these
hyperparameters and we obtain similar results for a range of different values.

\cref{eq:nmf-problem} is a well-known problem. We solve it with a
coordinate-descent algorithm described in \citet{cichocki2009fast} and
implemented in \verb|scikit-learn| \citep{pedregosa2011scikit}.
Then, let $\bm{N} \in \mathbb{R}^{d \times d}$
be the diagonal matrix containing the Euclidean norms of the columns of
$\bm{U}$, \ie such that
$n_{ii} = ||\bm{u}_{*,i}||_2$
and let
$\tilde{\bm{V}} = \bm{N} \, \bm{V}$.
We define the word similarity matrix
$\bm{A} = \tilde{\bm{V}}^T \, \tilde{\bm{V}} \in \mathbb{R}^{v \times v}$.
This matrix is a denoised, low-rank approximation of the corpus covariance.
Indeed,
\begin{IEEEeqnarray}{rCl}
  \bm{X}^T \, \bm{X}
  & \approx &
  \left( \bm{U}\, \bm{V} \right)^T \, \bm{U} \, \bm{V} \\
  & = &
 \bm{V}^T \, \bm{N}^T \, \left(\bm{U}\,\bm{N}^{-1}\right)^T \, \bm{U} \,
 \bm{N}^{-1} \, \bm{N} \, \bm{V} \\
 & \approx &
 \tilde{\bm{V}}^T \, \tilde{\bm{V}} \; .
\end{IEEEeqnarray}
The last approximation is justified by the fact that the columns of $\bm{U} \in
\mathbb{R}^{n \times d}$ are almost orthogonal, and $\bm{U}^T\,\bm{U}$ is almost
a diagonal matrix. This is what we observe in practice, and is due to the fact
that $n \approx 13\,000$ is much larger than $d = 300$, and that to minimize the
reconstruction error in \cref{eq:nmf-problem} the columns of $\bm{U}$ have an
incentive to span a large subspace of $\mathbb{R}^n$.

The similarity matrix $\bm{A}$ contains the inner products of the
low-dimensional embeddings of the terms in our vocabulary. We form the
matrix $\bm{T}$ by dividing the rows of $\bm{A}$ by their $\ell_1$
norm:
\begin{equation}
  \label{eq:transition-matrix-def}
  t_{i,j} = \frac{a_{i,j}}{||\,\bm{a}_{i}\,||_1}\;
  \forall \;
  i = 1 \dots v,\,j = 1 \dots v\, .
\end{equation}
This normalization ensures that terms that have many neighbors are not given
more importance in the smoothed representation.
The smoothing matrix that we use is then defined as:
\begin{equation}
\bm{S} = (1 - \alpha)\, \bm{I} + \alpha \, \bm{T} \; ,
\end{equation}
with $0 < \alpha < 1$ (in our experiments $\alpha$ is set to $0.1$).
This smoothing matrix is a mixture of the identity matrix and the term
associations $\bm{T}$. The model is not very sensitive to the parameter $\alpha$
as long as it is chosen small enough for terms actually present in the query to
have a higher weight than terms introduced by the query expansion. This prevents
degrading performance for documents which contain well-encoded terms, which
obtain good prediction even without smoothing.
This explains why in \cref{fig:overview}, the prediction for ``visual'' relies
mostly on the regression coefficient for this exact term, whereas the prediction
for ``agnosia'' relies on coefficients of terms that are \emph{related} to
``agnosia'' -- ``agnosia'' itself is not kept by the feature selection
procedure.

The smoothed representation for a query $\bm{q}$ becomes:
\begin{equation}
 \bm{x} = \bm{S}^T \bm{q} \in \mathbb{R}^v
\end{equation}
where
$\bm{q} \in \mathbb{R}^v$ is the \ac{TFIDF} representation of the query in large
vocabulary space, and
$\bm{S} \in \mathbb{R}^{v \times v}$ is the smoothing matrix.
And the prediction for $q$ is:
\begin{equation}
  \hat{\bm{y}} =  \hat{\B} \, \bm{P} \, \bm{S}^T \bm{q},
\end{equation}
where
$\bm{P} \in \mathbb{R}^{u \times v}$ is the projection onto the useful vocabulary
(selected features),
$\hat{\B} \in \mathbb{R}^{p \times u}$ are the estimated linear regression coefficients,
$\hat{\bm{y}} \in \mathbb{R}^{p}$ is the predicted map.

\section{Additional details on validation experiments}

\begin{figure}
  \iffinal
    \centerline{%
        \includegraphics[width=.9\textwidth]{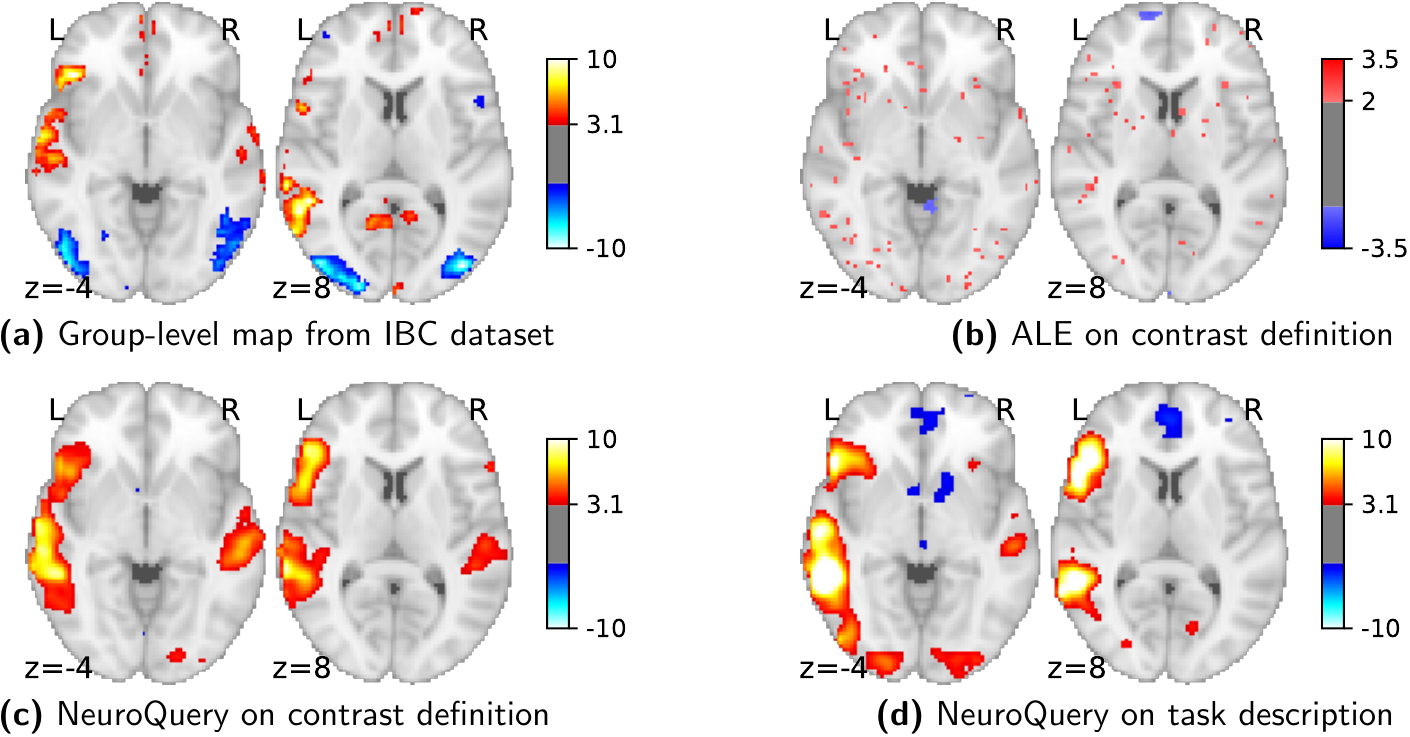}%
    }
  \fi%
          
          \caption{
            \textbf{Using meta-analysis to interpret \ac{fMRI} maps}. Example of the
            ``Read pseudo-words \vs consonant strings'' contrast, derived from
            the \ac{RSVP} language task in the \ac{IBC} dataset.
            \textbf{\sffamily (a)}: the
            group-level map obtained from the actual \ac{fMRI} data from \ac{IBC}.
            \textbf{\sffamily (b)}: \ac{ALE} map
            using the 29 studies in our corpus that contain all 5 terms from the
            contrast name.
            \textbf{\sffamily (c)}:
            NeuroQuery map obtained from the contrast name.
            \textbf{\sffamily (d)}:
            NeuroQuery map obtained from the page-long \ac{RSVP} task
            description in the \ac{IBC} dataset documentation:
            \url{https://project.inria.fr/IBC/files/2019/03/documentation.pdf}
          }\label{fig:neuroquery-experiments-ibc-roi-full-maps}
\end{figure}

\subsection{Example Meta-analysis results for the \ac{RSVP} language task from the
  \ac{IBC} dataset.}\label{section:details-ibc-experiment}

Here we provide more details on the meta-analyses for ``Read pseudo-words vs
consonant strings'' shown in \cref{fig:roi-ibc}.
the PMIDS of the studies included in the GingerALE meta-analysis are:
15961322, 16574082, 16968771, 17189619, 17884585, 17933023, 18272399, 18423780, 18476755, 18778780, 19396362, 19591947, 20035884, 20600985, 20650450, 20961169, 21767584, 22285025, 22659111, 23117157, 23270676, 24321558, 24508158, 24667455, 25566039, 26017384, 26188258, 26235228, 28780219.
Representing a total of 29 studies and 2025 peak activation coordinates.
They are the studies from our corpus (the largest existing corpus of text and
peak activation coordinates, with $\approx$ 14\,000 studies) which contain the
terms: ``reading'', ``pseudo'', ``word'', ``consonant'' and ``string''.
The map shown on the right of \cref{fig:roi-ibc} was obtained with GingerALE, 5000
permutations and the default settings otherwise. Note that an unrealistically
low threshold is used for the display because the map would be empty otherwise.
\cref{fig:neuroquery-experiments-ibc-roi-full-maps} displays more maps
with different analysis strategies: the details of the original contrasts
and the difference between running NeuroQuery the contrast definition or
the task definition. The task definition leads to predicted activations
in the early visual cortex, as in the actual group-level maps from the
experiment but unlike the predictions from the contrast definition, as
the later contains no information on the stimulus modality.

\subsection{NeuroQuery performance on unseen pairs of
terms}\label{section:unseen-pairs-performance}

\begin{figure}[t]
  \iffinal
    \includegraphics[width=\textwidth]{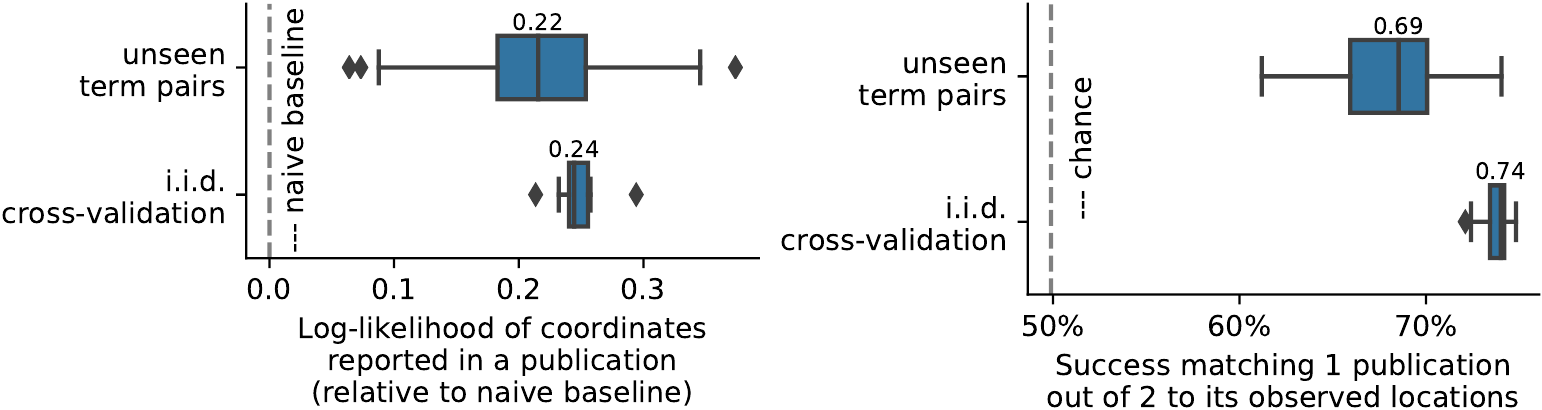}
  \fi%
  \caption{\textbf{Quantitative evaluations on unseen pairs} A
    quantitative comparison of prediction on random unseen studies
    (i.i.d. cross-validation) to
    prediction on studies containing pairs of terms never seen before,
    using the two measures of predictions performance 
    (as visible on \cref{fig:log-likelihood-box-plot} for standard
    cross-validation).
    }
  \label{fig:unseen_combination_prediction_performance}
\end{figure}

\cref{fig:unseen_combination} shows in a qualitative way that 
NeuroQuery can produce useful brain maps on a combination of terms
that have not been studied together. To give a quantitative evaluation
that is not limited to a specific pair of terms, we perform a systematic
experiment, studying prediction on many unseen pairs of term. For this
purpose, we chose pairs of terms in our full corpus and leave out all the
studies where both of these terms appear. We train a NeuroQuery model on
the reduced corpus of studies obtained by excluding studies with both
terms, and evaluate its predictions on the left-out studies.

We choose terms that appear simultaneously in studies frequently (more
than 500) to ensure a good estimation of the combined locations for these
terms in the test set, but not too frequently (less than 1000), to avoid
depleting the training set too much. Indeed, removing the studies for
both terms from the corpus not only decreases the statistical power to
map these terms but also, more importantly, it creates a negative
correlation between these terms. Out of these terms, we select 1000 out
random as a left-out and run the experiment 1000 times.

\begin{figure}[t]
\begin{minipage}{.495\linewidth}%
  \caption{\textbf{Consistency between prediction of unseen pairs and
    meta-analysis}
    The Pearson correlation between the map predicted by NeuroQuery on a
    pair of unseen terms and the average density of locations reported on
    the studies containing this pair of terms (hence excluded from the
    training set of NeuroQuery).}%
  \label{fig:unseen_combination_pearson}%
\end{minipage}%
\hfill%
\begin{minipage}{.5\linewidth}%
  \iffinal
    \includegraphics[width=\textwidth]{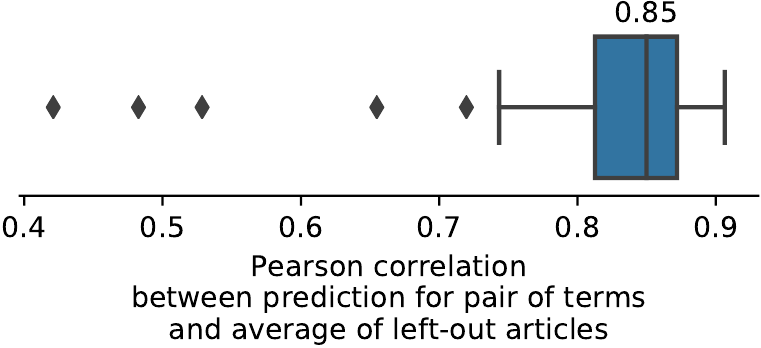}
  \fi%
\end{minipage}%
\end{figure}

To evaluate NeuroQuery's prediction on these unseen pairs of terms, we
first use the same metrics as in \cref{encoding-papers-experiment}.
\cref{fig:unseen_combination_prediction_performance}-left shows the
log-likelihood of coordinates reported in a publication evaluated on
left-out studies that contain the combination of terms excluded from the
train set. Compared to testing on a random subset of studied, identically
distributed to the training, 
there is a slight decrease in likelihood but it is small
compared to the variance between cross-validation runs.
\cref{fig:unseen_combination_prediction_performance}-right shows results
for our other validation metric \citep[adapted
from][]{mitchell2008predicting}: matching 1 publication out of 2 to its
observed locations. The decrease in performance is more
marked. However, it should be noted that the task is more difficult when
the test set is made only of publications that all contain two terms, as
these publications are all more similar to each other than random
publications from the general corpus.

To gauge the quality of the maps on unseen pairs, and not only how well
the corresponding publications are captured, 
\cref{fig:unseen_combination_pearson}
shows the Pearson correlation between the predicted brain map and the
average density of the reported locations in the left-out studies. The
excellent median Pearson correlation of .85 shows that the predicted
brain map is indeed true to what a meta-analysis of these studies would
reveal.

\subsection{NeuroQuery prediction performance without anatomical terms}\label{section:quantitative-results-details}

\begin{figure}[!htb]
  \iffinal
  \hspace{-50pt}
    \includegraphics[width=1.2\textwidth]{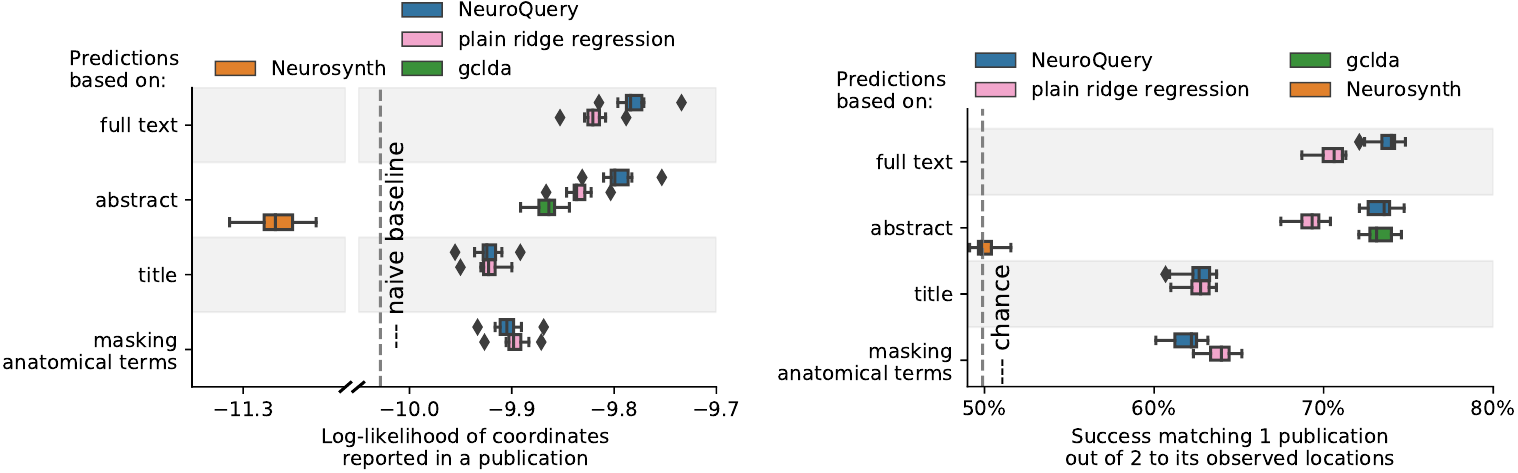}
  \fi%

  \begin{minipage}[t]{\textwidth}
  \caption{\textbf{Explaining coordinates reported in unseen
    studies.}
    left: log-likelihood of reported coordinates in test articles.
    right: how often the predicted map is closer to the true coordinates than to
    the coordinates for another article in the test set
    \citep{mitchell2008predicting}.
    The boxes represent the first, second and third quartiles of scores across 16
    cross-validation folds. Whiskers represent the rest of the distribution,
    except for outliers, defined as points beyond 1.5 times the \acs{IQR} past
    the low and high quartiles, and represented with diamond fliers.
  }
  \label{fig:log-likelihood-box-plot-detailed}
  \end{minipage}
\end{figure}

In \cref{fig:log-likelihood-box-plot-detailed}, we present an additional
quantitative measure of prediction performance.
We delete all terms that are related to anatomy in test
articles, to see how NeuroQuery performs without these highly predictive
features, which may be missing from queries related to brain function.
As the \ac{GCLDA} and NeuroSynth tools are designed to work with NeuroSynth
data, they are only tested on NeuroSynth's \ac{TFIDF} features, which represent
the articles' abstracts.



\subsection{Variable terminology}

\begin{figure}[!htb]
  \centering
  \iffinal
    \includegraphics[width=\textwidth]{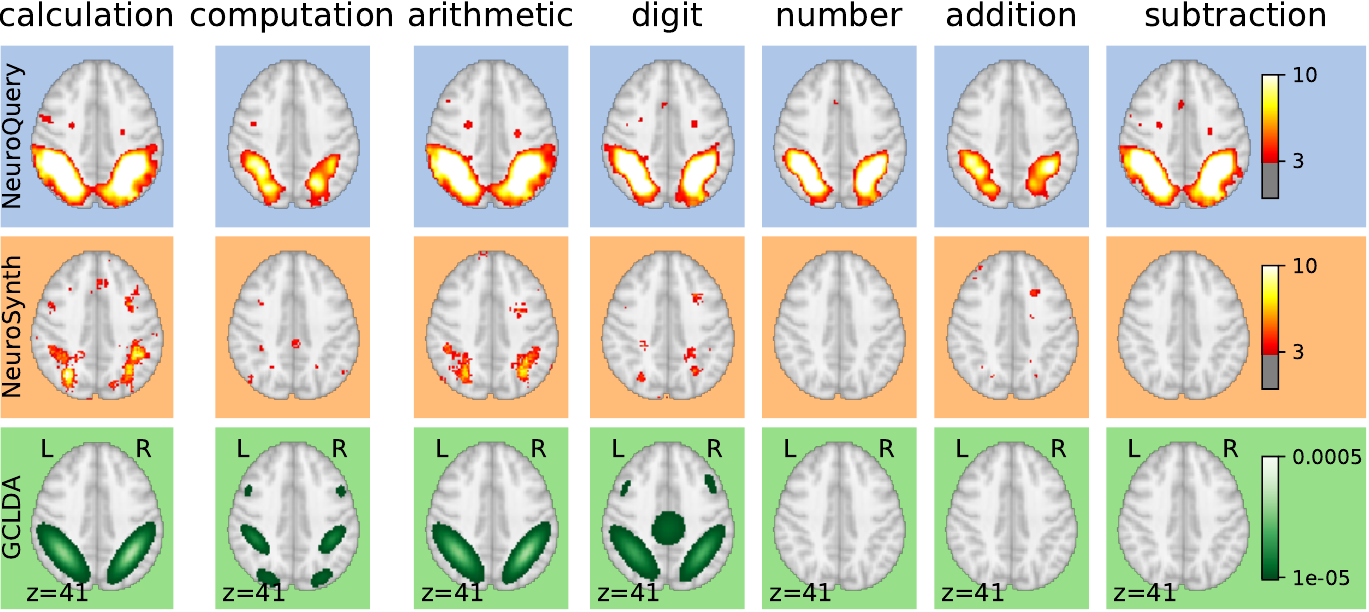}
  \fi%
  \caption{\textbf{Taming arbitrary query variability} Maps obtained
for a few words related to mental arithmetic. By correctly capturing the fact
that these words are related, NeuroQuery can use its map for easier words like
``calculation'' and ``arithmetic'' to encode terms like ``computation'' and
``addition'' that are difficult for meta-analysis.}
  \label{fig:maps-related-to-calculation-extended}
\end{figure}

In \cref{fig:maps-related-to-calculation-extended}, we show predictions for some
terms related to mental arithmetic. NeuroQuery's semantic smoothing produces
consistent results for related terms.

\subsection{Comparison with the BrainPedia \ac{IBMA} study}\label{subsection:comparison-with-brainpedia}

To compare maps produced by NeuroQuery with a reliable ground truth, we use the
BrainPedia study \citep{varoquaux2018atlases}, which exploits \ac{IBMA} to
produce maps for 19 cognitive concepts. Indeed, when it its feasible, \ac{IBMA}
of manually selected studies produces high-quality brain maps and has been used
as a reference for \ac{CBMA} methods \citep{salimi2009meta}.
We download the BrainPedia maps and their cognitive labels from the NeuroVault
platform\footnote{\url{https://neurovault.org/collections/4563/}}.
BrainPedia maps combine forward and reverse inference, and are thresholded to
identify regions that are both recruited and predictive of each cognitive
process. We treat these maps as a binary ground truth: above-threshold voxels
are relevant to the map's label.
For each label, we obtain a brain map from NeuroQuery, NeuroSynth and
\ac{GCLDA}. We compare these results to the BrainPedia thresholded maps and
measure the Area Under the \ac{ROC} Curve. This standard classification metric
measures the probability that a voxel that is active in the BrainPedia reference
map will be given a higher intensity in the NeuroQuery prediction than a voxel
that is inactive in the BrainPedia map.

\begin{figure}[h!]
  \iffinal
    \includegraphics[width=1\textwidth]{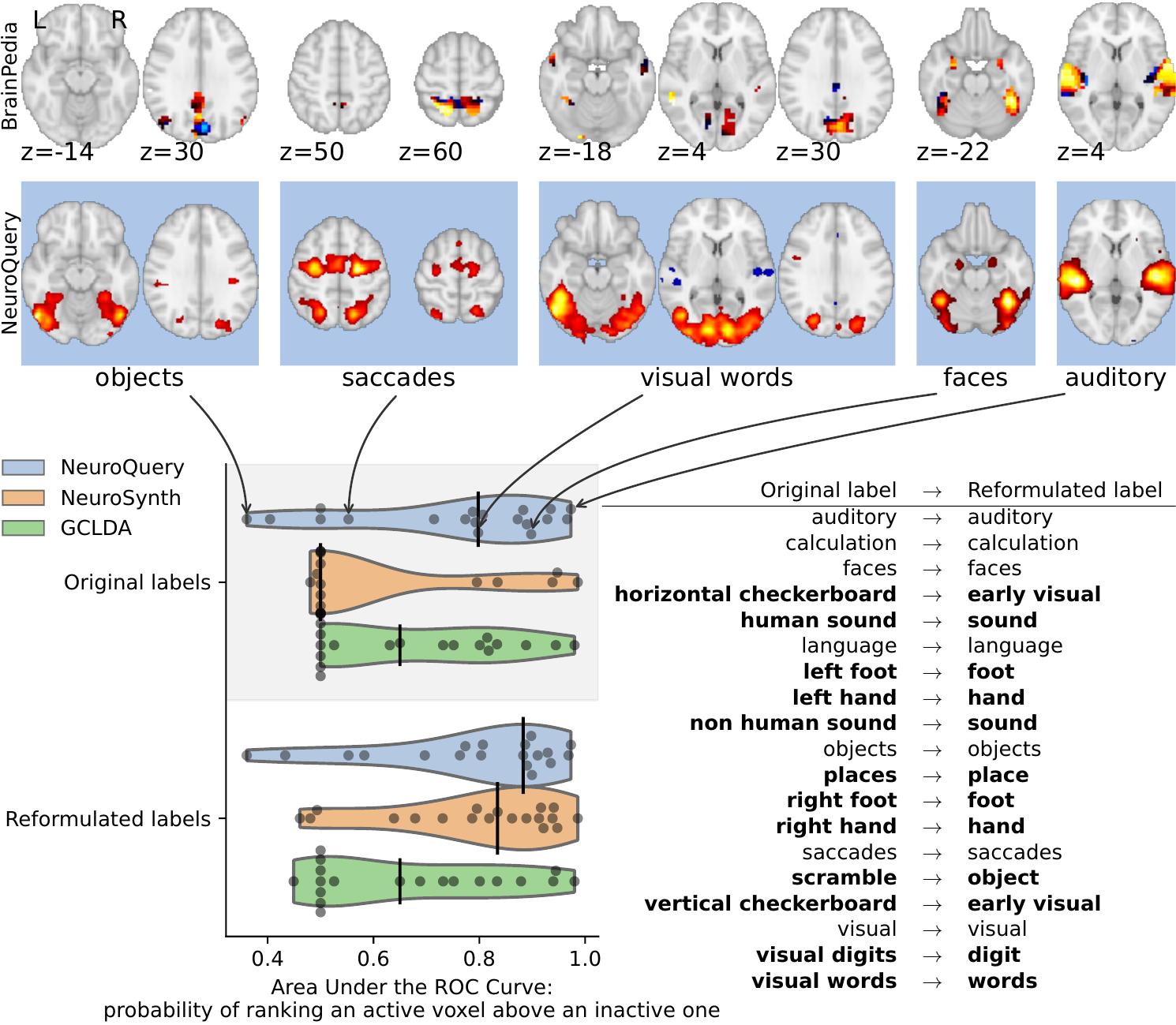}
  \fi%
  \caption{%
    \textbf{Comparison of \ac{CBMA} maps with \ac{IBMA} maps from the BrainPedia
    study.}
    We use labelled and thresholded maps resulting from a manual \ac{IBMA}.
    The labels are fed to NeuroQuery, NeuroSynth and \ac{GCLDA} and their
    results are compared to the reference by measuring the Area under the
    \ac{ROC} Curve. The black vertical bars show the median.
    When using the original BrainPedia labels, NeuroQuery performs relatively
    well but NeuroSynth fails to recognize most labels. When reformulating the
    labels, \ie replacing them with similar terms from NeuroSynth's vocabulary,
    both NeuroSynth and NeuroQuery match the manual \ac{IBMA} reference for most
    terms.
    On the top, we show the BrainPedia map (first row) and NeuroQuery prediction
    (second row) for the quartiles of the \ac{AUC} obtained by NeuroQuery on the
    original labels. A lower \ac{AUC} for some concepts can sometimes be
    explained by a more noisy BrainPedia reference map.
  }
  \label{fig:brainpedia-scores-with-maps}
\end{figure}

We consider two settings. First, we use the original labels provided in the
NeuroVault metadata. However, some of these labels are missing from the
NeuroSynth vocabulary. In a second experiment, we therefore replace these labels
with the most similar term we can find in the NeuroSynth vocabulary. These
replacements are shown in \cref{fig:brainpedia-scores-with-maps}.

When replacing the original labels with less specific terms understood by
NeuroSynth, both NeuroQuery and NeuroSynth perform well: NeuroQuery's median
\ac{AUC} is 0.9 and NeuroSynth's is 0.8.
When using the original labels, NeuroSynth fails to produce results for many
labels as they are missing from its vocabulary. NeuroQuery still performs well
on these uncurated labels with a median \ac{AUC} of 0.8.
Finally, we can note that although the BrainPedia maps come from \ac{IBMA}
conducted on carefully selected \ac{fMRI} studies, they also contain some noise.
As can be seen in \cref{fig:brainpedia-scores-with-maps}, BrainPedia maps that
qualitatively match the domain knowledge also tend to be close to the \ac{CBMA}
results produced by NeuroQuery and NeuroSynth.

\subsection{Comparison with Harvard-Oxford anatomical atlas}\label{subsection:harvard-oxford-experiment}

\begin{figure}[h!]
  \iffinal
    \includegraphics[width=1\textwidth]{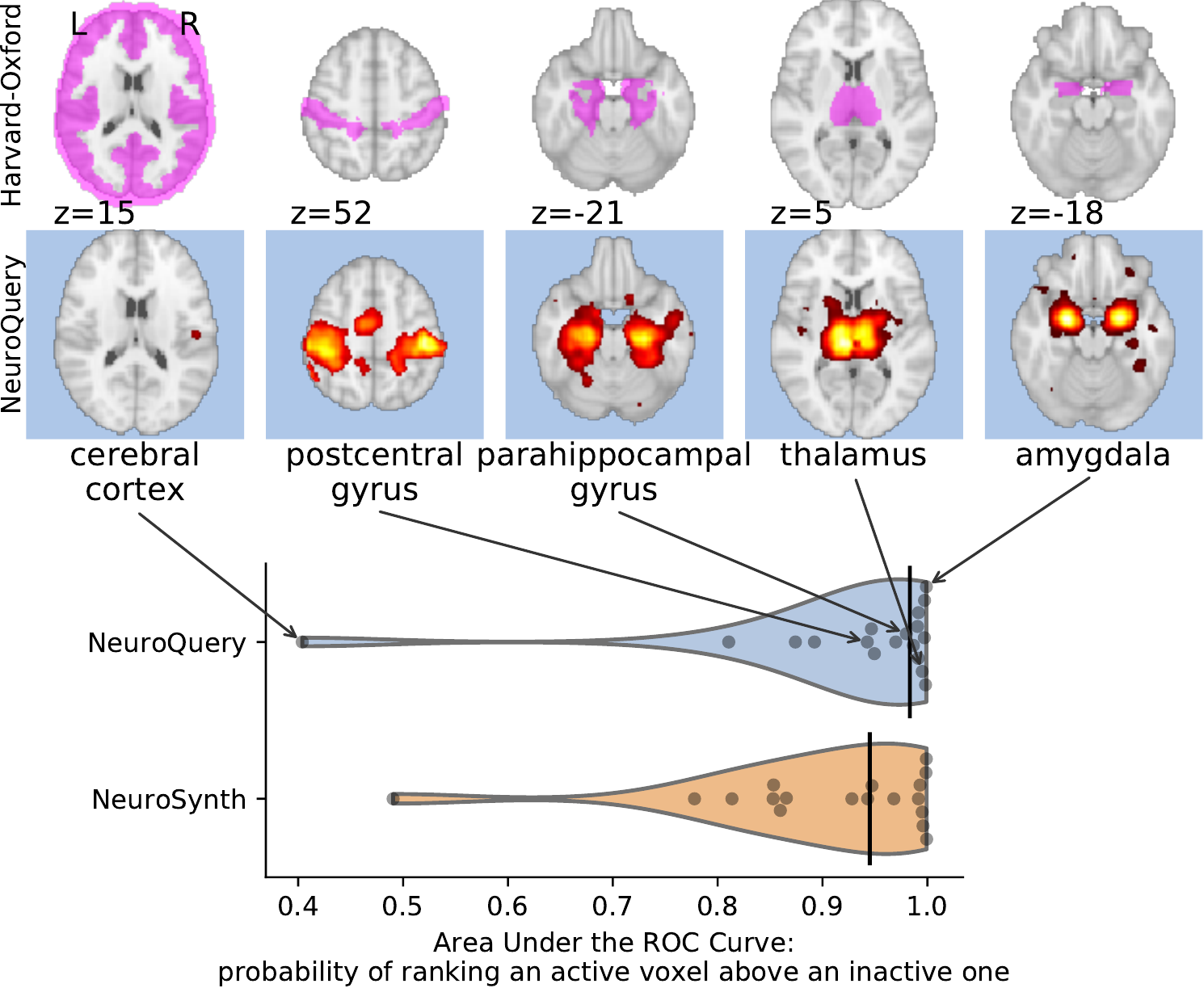}
  \fi%
  \caption{
    \textbf{Comparison of predictions with regions of the Harvard-Oxford anatomical
    atlas.}
    Labels of the Harvard-Oxford anatomical atlas present in NeuroSynth's
    vocabulary are fed to NeuroSynth and NeuroQuery. The meta-analytic maps are
    compared to the manually segmented reference by measuring the Area Under the
    \ac{ROC} Curve. The black vertical bars show the median.
    Both NeuroSynth and NeuroQuery achieve a median \ac{AUC} above 0.9.
    On the top, we show the atlas region (first row) and NeuroQuery prediction
    (second row) for the quartiles of the NeuroQuery \ac{AUC} scores.
  }
  \label{fig:harvard-oxford}
\end{figure}

Here, we compare \ac{CBMA} maps to manually segmented regions of the
Harvard-Oxford anatomical atlas \citep{desikan2006automated}. We feed the labels
from this atlas to NeuroQuery, NeuroSynth and \ac{GCLDA} and compare the
resulting maps to the atlas regions.
This experiment provides a sanity check that relies on an excellent ground
truth, as the atlas regions are labelled and segmented by experts.
For simplicity, atlas labels absent from NeuroSynth's vocabulary are discarded.
For the remaining 18 labels, we compute the Area Under the \ac{ROC} Curve of the
maps produced by each meta-analytic tool.
This experiment is therefore identical to the one presented in
\cref{subsection:comparison-with-brainpedia}, except that the reference ground
truth is a manually segmented anatomical atlas, and that we do not consider
reformulating the labels. \ac{GCLDA} is not used in this experiment as the
trained model distributed by the authors does not recognize anatomical terms.
We observe that both NeuroSynth and NeuroQuery match closely the reference
atlas, with a median \ac{AUC} above 0.9, as seen in \cref{fig:harvard-oxford}.

\subsection{Comparison with NeuroSynth on terms with strong activations}\label{subsection:comparison-with-neurosynth}
\begin{figure}
  \iffinal
    \centerline{%
    \includegraphics[width=.8\textwidth]{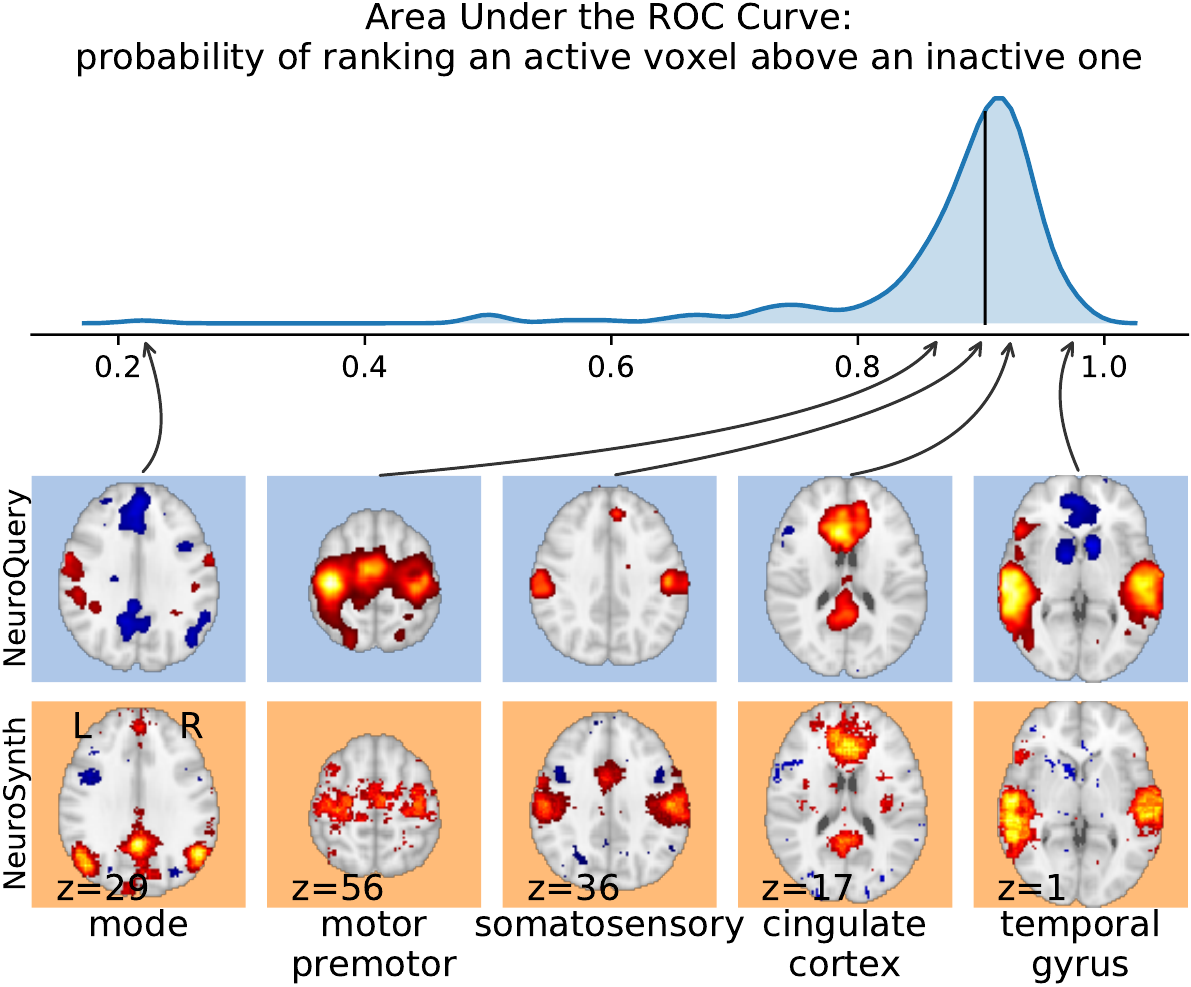}%
    }
  \fi%
  \caption{\textbf{Comparison with NeuroSynth.} NeuroSynth maps are thresholded
    controlling the \acs{FDR} at 1\%. The 200 words with the largest number of
    active voxels are selected and NeuroQuery predictions are compared to the
    NeuroSynth activations by computing the Area Under the ROC Curve. The
    distribution of the AUC is shown on the top. The
    vertical black line shows the median (0.90). On the bottom, we show the
    NeuroQuery maps (first row) and NeuroSynth activations (second row) for the
    quartiles of the NeuroQuery AUC scores.}
  \label{fig:neuroquery-comparison-with-neurosynth}
\end{figure}

As NeuroSynth performs a statistical test, when a term has a strong link with
brain activity and is popular enough for NeuroSynth to detect many activations,
the resulting map is trustworthy and can be used as a reference.
Moreover, it is a well-established tool that has been adopted by the
neuroimaging community.
Here, we verify that when a term is well captured by NeuroSynth, NeuroQuery
predicts a similar brain map.
To identify terms that NeuroSynth captures well, we compute the NeuroSynth maps
for all the terms in NeuroSynth's vocabulary. We use the Benjamini-Hochberg
procedure to threshold the maps, controlling the \ac{FDR} at 1\%. We then select
the 200 maps with the largest number of active (above-threshold) voxels.
We use these activation maps as a reference to which we compare the NeuroQuery
prediction. For each term, we compute the Area Under the \ac{ROC} Curve: the
probability that a voxel that is active in the NeuroSynth map will have a higher
value in the NeuroQuery prediction than an inactive voxel.
We find that NeuroQuery and NeuroSynth's maps coincide well, with a median
\ac{AUC} of 0.90. The distribution of the \ac{AUC} and the brain map
corresponding to each quartile are shown in
\cref{fig:neuroquery-comparison-with-neurosynth}.

\section{Word occurrence frequencies across the corpus}\label{document-frequencies}

\begin{figure}
  \iffinal
    \centerline{\includegraphics[width=.8\textwidth]{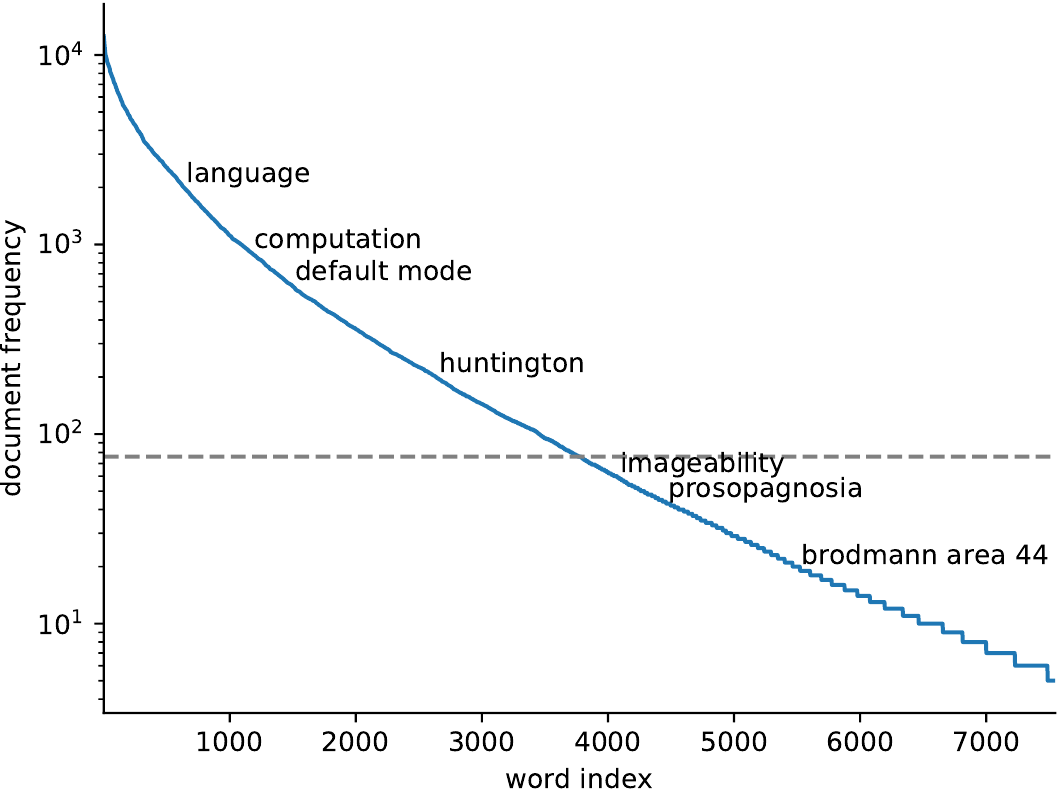}}%
  \fi%
  \caption{\textbf{Most terms occur in few documents} Plot of the document
    frequencies for terms in the vocabulary, sorted in decreasing order. While
    some terms are very frequent, occurring in over 12\,000 articles, most are
    very rare: half occur in less than 76 (out of 14\,000) articles.}
  \label{fig:zipf-law-appendix}

    \bigskip
    \begin{minipage}{.45\linewidth}
  \caption{\textbf{Benefit of using full-text articles}.
    Document frequencies (number of documents in which a word appears)
    for terms from the NeuroSynth vocabulary, in the NeuroSynth corpus ($x$
    axis) and the NeuroQuery corpus ($y$ axis).
    Words appear in much fewer documents in the NeuroSynth corpus
    because it only contains abstracts. Even when considering only terms present
    in the NeuroSynth vocabulary, the NeuroQuery corpus contains over 3M term-study
    associations -- 4.6 times more than NeuroSynth.
    }\label{fig:scatter-doc-frequencies-neuroquery-vs-neurosynth}
  \end{minipage}%
  \hfill
    \begin{minipage}{.43\linewidth}
  \iffinal
    \includegraphics[width=1.2\textwidth]{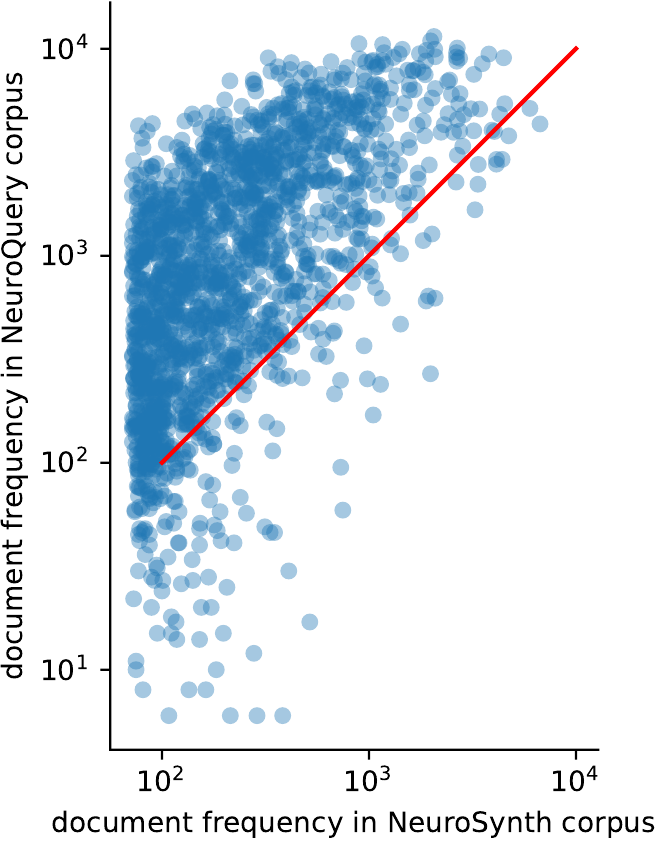}
  \fi%
    \end{minipage}%
\end{figure}

\begin{figure}
    \begin{minipage}{.2\linewidth}\sloppy
  \caption{Document frequencies for some example words, in NeuroQuery's and
    NeuroSynth's corpora.}
  \label{fig:example-document-frequencies}
    \end{minipage}%
    \begin{minipage}{.8\linewidth}
    \iffinal
      \includegraphics[width=\textwidth]{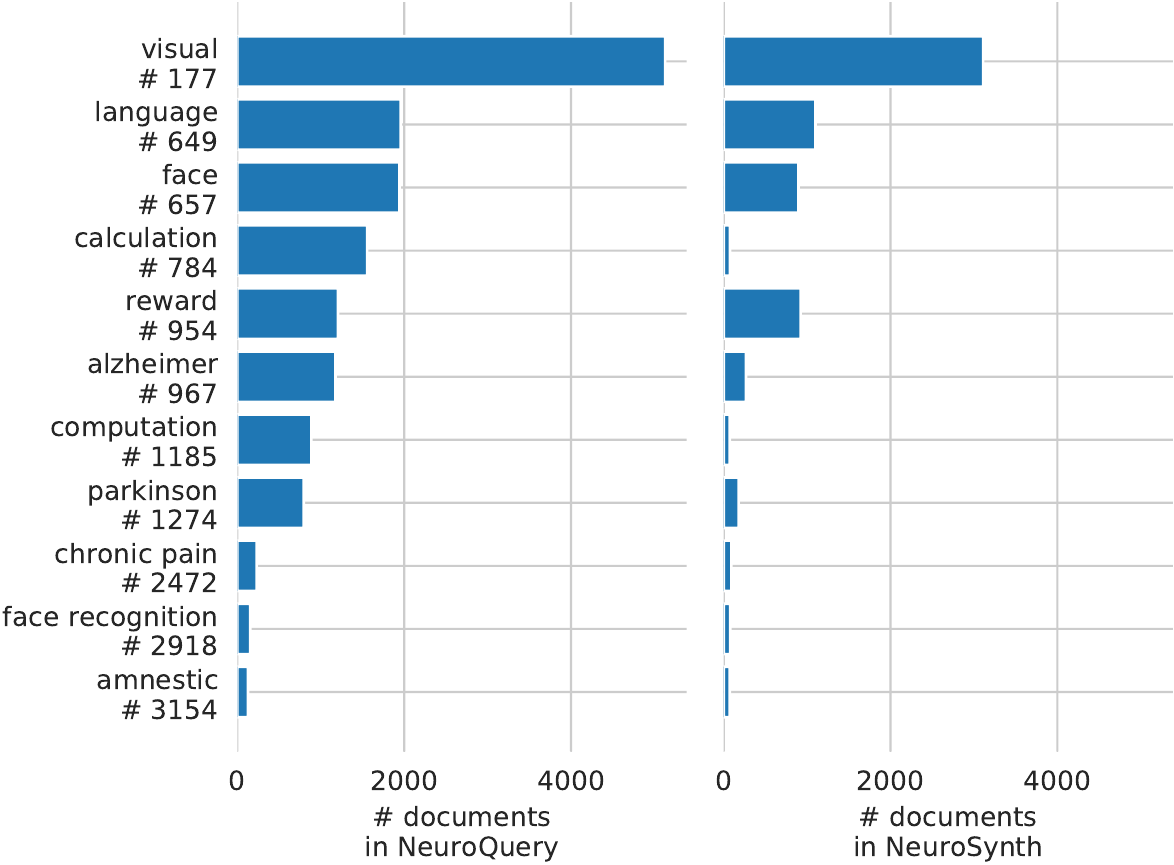}
    \fi%
    \end{minipage}%
\end{figure}

As shown on \cref{fig:zipf-law-appendix}, most words occur in very few documents, which
is why correctly mapping rare words is important.
The problem of rare words is more severe in the NeuroSynth corpus, which
contains only the abstracts. As the NeuroQuery corpus contains the full text of
the articles (around 20 times more text), more occurrences of a unique term in a
document are observed, as shown in
\cref{fig:scatter-doc-frequencies-neuroquery-vs-neurosynth}, and in
\cref{fig:example-document-frequencies} for a few example terms.
\begin{figure}
    \begin{minipage}{.4\linewidth}%
  \caption{How often a phrase from the vocabulary (e.g. ``face recognition'')
    occurs, versus at least one of its constituent words (e.g. ``face'').
    Expressions involving several words are typically very rare.}
  \label{fig:intersection-power-failure}
    \end{minipage}%
    \begin{minipage}{.6\linewidth}%
  \iffinal
    \includegraphics[width=\textwidth]{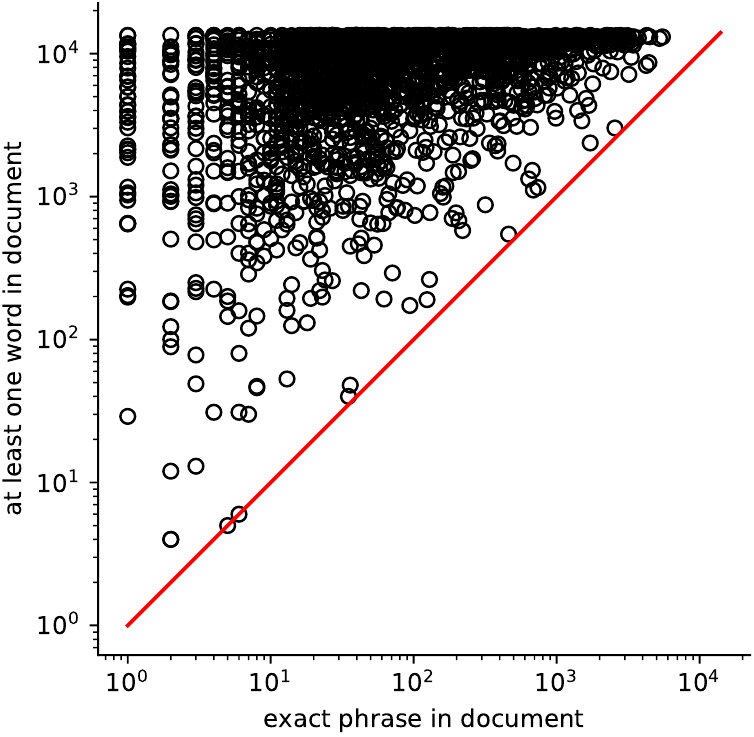}
  \fi%
    \end{minipage}%
\end{figure}
%
\para
Document set intersections lack statistical power. For example, ``face
perception'' occurs in 413 articles, and ``dementia'' in 1312. 1703 articles
contain at least one of these words and could be used for a multivariate
regression's prediction for the query ``face perception and dementia''. Indeed,
denoting $c$ the dual coefficients of the ridge regression and $\X$ the training
design matrix, the prediction for a query $q$ is $q^t \X^tc$, and any document
that has a nonzero dot product with the query can participate in the prediction.
However, only 22 documents contain both terms and would be used
with the classical meta-analysis selection, which would lack statistical power
and fail to produce meaningful results.
Exact matches of multi-word expressions such as ``creative problem solving'', ``
facial trustworthiness recognition '', ``positive feedback processing'',
``potential monetary reward'', ``visual word recognition'' (all cognitive atlas
concepts, all occurring in less than 5 / 10\,000 full-text articles), are very
rare -- and classical meta-analysis thus cannot produce results for such
expressions. In \cref{fig:intersection-power-failure}, we compare the frequency
of multi-word expressions from our vocabulary (such as ``face recognition'')
with the frequency of their constituent words. Being able to combine words in an
additive fashion is crucial to encode such expressions into brain space.

\section{Details on choice of vocabulary}
\subsection{Details on the Medical Subject Headings}\label{parts-of-mesh-used}

The \acf{MeSH} are concerned with all of medicine. We only included in
NeuroQuery's vocabulary the parts of this graph that are relevant for
neuroscience and psychology. Here we list the branches of \acf{MeSH} that we
included in our vocabulary:


Neuroanatomy: 'A08.186.211'

Neurological disorders: 'C10.114', 'C10.177', 'C10.228', 'C10.281', 'C10.292',
'C10.314', 'C10.500', 'C10.551', 'C10.562', 'C10.574', 'C10.597', 'C10.668',
'C10.720', 'C10.803', 'C10.886', 'C10.900'

Psychology: 'F02.463', 'F02.830', 'F03', 'F01.058', 'F01.100', 'F01.145',
'F01.318', 'F01.393', 'F01.470', 'F01.510', 'F01.525', 'F01.590', 'F01.658',
'F01.700', 'F01.752', 'F01.829', 'F01.914'

Many MeSH terms are too rare to be part of NeuroQuery's vocabulary. Some are too
specific, \eg ``Diffuse Neurofibrillary Tangles with Calcification''. More
importantly, many terms are absent because for each heading, MeSH provides many
\emph{Entry Terms} -- various ways to refer to a concept, some of which are
almost never used in practice in the text of publications. For example
NeuroQuery recognizes the MeSH \emph{Preferred Term} ``Frontotemporal Dementia''
but not some of its variations
(\url{https://meshb.nlm.nih.gov/record/ui?ui=D057180}) such as ``Dementia,
Frontotemporal'', ``Disinhibition-Dementia-Parkinsonism-Amyotrophy Complex'', or
``HDDD1''. Note that even when absent from the vocabulary as single phrases,
many of these variations can be parsed as a combination of several terms,
resulting in a similar brain map as the one obtained for the preferred term.

\subsection{Atlas labels included in the vocabulary}\label{atlases-used}

The labels from the 12 atlases shown in \cref{table:atlases-used}
were included in the NeuroQuery vocabulary.

\begin{table}[h!]
\begin{tabular}{p{.2\textwidth}p{.8\textwidth}}
                       name &                                                                        url \\
  \hline \\
                  talairach &                                     \url{http://www.talairach.org/talairach.nii} \\
             harvard\_oxford &               \url{http://www.nitrc.org/frs/download.php/7700/HarvardOxford.tgz} \\
                  destrieux &              \url{ https://www.nitrc.org/frs/download.php/7739/destrieux2009.tgz } \\
                        aal &                                             \url{ http://www.gin.cnrs.fr/AAL-217 } \\
                 JHU-labels &                  \url{ https://fsl.fmrib.ox.ac.uk/fsl/fslwiki/Atlases\#JHU-labels } \\
        Striatum-Structural &         \url{ https://fsl.fmrib.ox.ac.uk/fsl/fslwiki/Atlases\#Striatum-Structural } \\
                        STN &                         \url{ https://fsl.fmrib.ox.ac.uk/fsl/fslwiki/Atlases\#STN } \\
 Striatum-Connectivity-7sub &  \url{ https://fsl.fmrib.ox.ac.uk/fsl/fslwiki/Atlases\#Striatum-Connectivity-7sub } \\
                    Juelich &                     \url{ https://fsl.fmrib.ox.ac.uk/fsl/fslwiki/Atlases\#Juelich } \\
                        MNI &                         \url{ https://fsl.fmrib.ox.ac.uk/fsl/fslwiki/Atlases\#MNI } \\
                 JHU-tracts &                  \url{ https://fsl.fmrib.ox.ac.uk/fsl/fslwiki/Atlases\#JHU-tracts } \\
                   Thalamus &                    \url{ https://fsl.fmrib.ox.ac.uk/fsl/fslwiki/Atlases\#Thalamus } \\
\end{tabular}
  \caption{Atlases included in NeuroQuery's vocabulary}\label{table:atlases-used}
\end{table}

\section{NeuroSynth posterior probability maps}\label{ns-posterior-proba-maps}

The NeuroSynth maps shown in \cref{fig:coef-consistency} are the NeuroSynth
``association test'' maps. For completeness, here we show the other kind of map
that NeuroSynth can produce, called ``posterior probability'' maps.

\begin{figure}[hb]
  \iffinal
    \includegraphics[width=\textwidth]{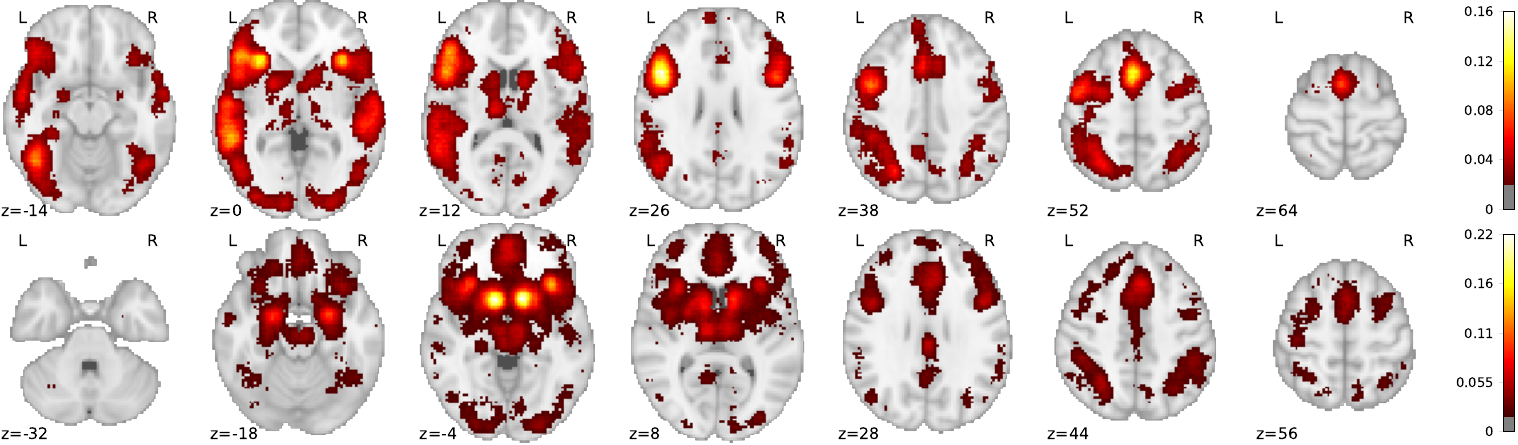}
  \fi%
  \caption{NeuroSynth posterior probability maps for ``language'' (top) and
    ``reward'' (bottom), using the full corpus.}
  \label{fig:neurosynth-posterior-proba-maps}
\end{figure}

\end{document}